\documentclass[12pt]{article}
\usepackage{latexsym}
\usepackage{epsfig,amssymb,euscript, mathrsfs}
\usepackage{amsmath}
\newsavebox{\ns}
\newsavebox{\dbrane}
\newsavebox{\dbshort}

\def\be{\begin{equation}}
\def\ee{\end{equation}}
\def\bea{\begin{eqnarray}}
\def\eea{\end{eqnarray}}

\def\be{\begin{equation}}
\def\ee{\end{equation}}
\def\ba{\begin{eqnarray}}
\def\ea{\end{eqnarray}}

\newcommand{\nn}{\nonumber}

\def\Dslash{\,\,{\raise.15ex\hbox{/}\mkern-12mu D}}
\def\Dbarslash{\,\,{\raise.15ex\hbox{/}\mkern-12mu {\bar D}}}
\def\delslash{\,\,{\raise.15ex\hbox{/}\mkern-9mu \partial}}
\def\delbarslash{\,\,{\raise.15ex\hbox{/}\mkern-9mu {\bar\partial}}}
\def\pslash{\,\,{\raise.15ex\hbox{/}\mkern-9mu p}}
\def\calDslash{\,\,{\raise.15ex\hbox{/}\mkern-12mu {\cal D}}}

\newcommand\R{\mathbb{R}}
\newcommand\Z{\mathbb{Z}}

\newcommand\C{\mathbb{C}}

\newcommand\diff{\mathrm{d}}

\newcommand{\ii}{\mathrm{i}}

\newcommand{\gauge}{G}
\newcommand{\ex}{\mathrm{e}}
\newcommand{\zflip}{\mathbb{Z}_2^\mathrm{flip}}
\newcommand{\Da}{\Delta_{\mathrm{Ad}}}
\newcommand{\Db}{\Delta_{\mathrm{Bi}}}
\newcommand{\Phib}{\Phi_{\mathrm{Bi}}}
\newcommand{\Phia}{\Phi_{\mathrm{Ad}}}
\newcommand{\sgn}{\mathrm{sgn}}
\newcommand{\sumItoJ}{\sum_{\mathrm{fixed} \, I\rightarrow J}}
\newcommand{\sumJtoI}{\sum_{\mathrm{fixed} \ I\leftarrow J}}
\newcommand{\sumarrows}{\sum_{\mathrm{all}\ I\rightarrow J}}
\newcommand{\sumIfromtoJ}{\sum_{\mathrm{fixed} \ I\leftrightarrow J}}

\numberwithin{equation}{section}       % equation numbers in each section

\newcommand{\qed}{\nobreak \ifvmode \relax \else \ifdim\lastskip<1.5em \hskip-\lastskip \hskip1.5em plus0em minus0.5em \fi \nobreak \vrule height0.75em width0.5em depth0.25em\fi}

\begin{document}
\begin{titlepage}
\begin{center}
\today

\vskip 2 cm {\Large \bf The large $N$ limit of quiver matrix models\\}
\vskip 0.3 cm {\Large \bf and Sasaki-Einstein manifolds\\} 
\vskip 1.8 cm
{Dario Martelli$^1$ and James Sparks$^2$\\}
\vskip 1cm
$^1$\textit{Department of Mathematics, King's College, London, \\
The Strand, London WC2R 2LS,  United Kingdom\\}
\vskip 0.8cm
$^2$\textit{Mathematical Institute, University of Oxford,\\
24-29 St Giles', Oxford OX1 3LB, United Kingdom\\}
\vskip 1cm

\end{center}

\vskip 3 cm
\begin{abstract}
\noindent We study the matrix models that result from localization of the partition functions 
of $\mathcal{N}=2$ Chern-Simons-matter theories on the three-sphere. A large class of such 
theories are conjectured to be holographically dual to M-theory on Sasaki-Einstein seven-manifolds. 
We study the M-theory limit (large $N$ and fixed Chern-Simons levels) of these matrix models 
for various examples, and show that in this limit the free energy reproduces the 
expected AdS/CFT result of $N^{3/2}/\mathrm{Vol}(Y)^{1/2}$, where  $\mathrm{Vol}(Y)$ is the volume 
of the corresponding Sasaki-Einstein metric. More generally 
we conjecture a relation between the large $N$ limit of the partition function, interpreted as a 
function of trial R-charges, and the volumes of Sasakian metrics on links of Calabi-Yau four-fold 
singularities. We verify this conjecture for a family of $U(N)^2$ Chern-Simons quivers 
based on M2 branes at hypersurface singularities, and for a $U(N)^3$ theory based on M2 branes at a toric singularity.

\end{abstract}

\end{titlepage}
\pagestyle{plain}
\setcounter{page}{1}
\newcounter{bean}
\baselineskip18pt
\tableofcontents

%%%%%%%%%%%%%%%%%%%%%%%%%%%%%%%%%%%%%%%%%%%%%%%%%%%%%%%%%%%%%

\section{Introduction}

For a long time the low energy theory on multiple M2 branes had remained rather mysterious, and consequently the 
AdS$_4$/CFT$_3$ correspondence poorly understood. A breakthrough occurred
with the construction by Bagger-Lambert \cite{BL} and Gustavsson \cite{Gustavsson:2007vu}
of new maximally supersymmetric Chern-Simons-matter theories, which they proposed to describe the low energy limit of M2 branes. Inspired by these results,  
subsequently Aharony {\it et al} (ABJM) \cite{ABJM}  conjectured the equivalence of a certain ${\cal N}=6$ supersymmetric Chern-Simons-matter theory with gauge group $U(N)\times U(N)$ and Chern-Simons levels $(k,-k)$ with the M-theory backgrounds AdS$_4\times S^7/\Z_k$. In particular, the ABJM theory is believed to describe $N$ M2 branes at a $\C^4/\Z_k$ orbifold singularity. 
Despite this improved  understanding of the microscopic theory of multiple M2 branes, 
a field theory derivation of  the  famous $N^{3/2}$ 
scaling of the number of degrees of freedom \cite{Klebanov:1996un} on multiple M2 branes, in the  large $N$ limit, had
 remained elusive. Remarkably, in \cite{Drukker:2010nc} this AdS/CFT prediction has been confirmed by a purely field theoretic calculation in the ABJM model. 
 The large $N$ limit of a BPS Wilson loop in this model was also computed in \cite{Suyama:2009pd}.\footnote{We would like to thank Takao Suyama 
for pointing out reference \cite{Suyama:2009pd}.}  
  More recently in \cite{Herzog:2010hf} and \cite{Santamaria:2010dm} similar results have been obtained for classes of ${\cal N}=3$
Chern-Simons-quivers  \cite{Imamura:2008nn} with  M-theory duals of the form AdS$_4\times Y$, where $Y$ are tri-Sasakian manifolds \cite{Jafferis:2008qz}.  The key to these results is the computation of  \cite{Kapustin:2009kz}, showing that the path integral of a (Euclidean) superconformal field theory on $S^3$ \emph{localizes} to a matrix integral.  As we will review momentarily, 
the results of  \cite{Kapustin:2009kz} are  effectively applicable to theories with ${\cal N}\geq 3$ supersymmetry. 
 
 ${\cal N}=2$ supersymmetric field theories in three dimensions are  expected to share certain properties with ${\cal N}=1$ supersymmetric field theories in four dimensions, since 
 the number of supercharges is the same.  For the latter theories, the exact NSVZ beta functions and $a$-maximization \cite{Intriligator:2003jj} provide stringent constraints on the R-symmetry of superconformal theories, and indeed allow one to unambiguosly determine  R-charges and related trace anomaly coefficients in most cases. Vanishing of beta functions and $a$-maximization may also be used as diagnostic tests of the existence of strongly coupled superconformal fixed points. In the context of the AdS$_5$/CFT$_4$ correspondence, 
$a$-maximization has a geometric counterpart in the volume minimization of Sasaki-Einstein manifolds\footnote{See \cite{Gabella:2009wu, Gabella:2010cy} for extensions of these results to more general AdS$_5$ geometries of Type IIB supergravity.} \cite{Martelli:2005tp, Martelli:2006yb, Gauntlett:2006vf}. 
Intriguingly, it has been shown that the trial  $a$ function and the reciprocal of the trial volume 
function  are equal, even before being extremized \cite{Butti:2005vn, Eager:2010yu}.  The geometric results of \cite{Martelli:2005tp, Martelli:2006yb, Gauntlett:2006vf} hold in any dimension, and hence crucially also in seven dimensions. Via the AdS/CFT correspondence, this suggests that  a large class of ${\cal N}=2$ Chern-Simons-matter theories, characterized by having AdS$_4$ gravity duals, should in many ways behave similarly to ${\cal N}=1$ superconformal field theories in four dimensions. However, until now this has remained only wishful thinking. 

Very recently Jafferis \cite{Jafferis:2010un} (see also \cite{Hama:2010av}) has
extended the results of \cite{Kapustin:2009kz} by
 showing that  the partition function $Z$ of a general ${\cal N}=2$ supersymmetric field theory on $S^3$ reduces to a matrix integral. 
Putting the theory on the three-sphere, in a manner which preserves supersymmetry, requires the introduction of additional 
couplings between the matter fields and the curvature of $S^3$. These couplings are determined by a choice of 
R-symmetry. One may now regard these as \emph{trial} R-charges for a putative superconformal fixed point. 
Furthermore, \cite{Jafferis:2010un} conjectured that the 
R-charges of the matter fields at the  superconformal fixed point are determined by extremizing (the modulus of) the partition function. If correct, this proposal gives support to the idea that $Z$ plays a similar role, for ${\cal N}=2$ field theories in three dimensions, to the central charge function $a$ in ${\cal N}=1$ field theories in four dimensions. In this paper we will initiate an investigation of this idea by evaluating the relevant matrix integrals, in the large $N$ limit, in some ${\cal N}=2$ models. 
In particular, we will find supporting evidence for the following:

\vskip 4mm 

\noindent
{\bf Conjecture}:  For ${\cal N}=2$ Chern-Simons-matter gauge  theories with candidate Sasaki-Einstein gravity duals, in the large $N$ limit 
 the leading contribution to the free energy,  as a function of trial  R-charges $R_a$, is related to the Sasakian volume Vol$(Y)[\xi]$
 as a function of  the Reeb vector field $\xi$ via the formula
\bea\label{conjecture}
\lim_{N\to \infty}   - \log \left| Z [R_a] \right|  & = &  N^{3/2}\sqrt{\frac{2\pi^6}{27\, \mathrm{Vol}(Y)[\xi]}}~.
\eea

\vskip 4mm 

Here by ``candidate Sasaki-Einstein dual'' we mean that the field theory (at least) possesses a branch of the Abelian vacuum moduli space that is an isolated  Calabi-Yau four-fold conical singularity \cite{Martelli:2008si,Hanany:2008cd}. We refer the reader to \cite{Martelli:2005tp,Martelli:2006yb} for a detailed explanation of the volume function Vol$(Y)[\xi]$. 
In (\ref{conjecture}) the trial R-charges should in general be understood to be functions of the Reeb vector field $\xi$, as in \cite{Butti:2005vn}.
The method that we will use for computing the leading free energy in the large $N$ limit follows very closely 
 \cite{Herzog:2010hf}; in the latter reference this method was applied to compute the same quantity for ${\cal N}=3$ matrix models,
 of the type studied in \cite{Kapustin:2009kz}. 
  
 The rest of the paper is organized as follows. In section \ref{sec2} we review relevant aspects of $\mathcal{N}=2$ Chern-Simons-matter theories, 
 localization of the partition functions of such theories on $S^3$ to matrix integrals, and  the expected 
AdS/CFT dual gravitational results  in M-theory. In section \ref{gensection} we discuss the M-theory large $N$ 
limit of the partition functions of such theories, following similar methods to \cite{Suyama:2009pd, Herzog:2010hf}. In particular, 
we derive general formulas for the leading one-loop contributions for non-chiral theories. Section 
\ref{Ansec} contains an explicit verification of (\ref{conjecture}) for a family of 
examples describing M2 branes at certain hypersurface singularities   
  \cite{Martelli:2009ga}; these 
are in some sense a simple $\mathcal{N}=2$ generalization of the ABJM theory.   In section 
\ref{SPPsec} we verify the matching of functions in (\ref{conjecture}) for 
  a three node theory related to the suspended pinch point (SPP) singularity \cite{Hanany:2008fj}.
In section \ref{discus}  we briefly outline  directions for future work. In Appendix \ref{expansions}
we collect some series expansions used in the main text. Appendix 
\ref{interappe} derives a technical result on the range of validity of a Fourier series used in 
section \ref{SPPsec}.
  
  \vskip 3mm 
  
  \noindent
  {\bf Note}: As this paper was being completed we learned of the paper \cite{Nakwoo}, which we understand has 
  overlap with the results presented here.
  
\section{Localization of $\mathcal{N}=2$ Chern-Simons-matter theories on $S^3$}\label{sec2}

\subsection{$\mathcal{N}=2$ Chern-Simons-matter theories}\label{CSmatter}

We begin by reviewing the $\mathcal{N}=2$ Chern-Simons-matter theories of interest, following \cite{Martelli:2008si}. Earlier foundational work on this topic includes \cite{Schwarz:2004yj, ABJM, Gaiotto:2007qi}. 

A three-dimensional ${\cal N}=2$ vector multiplet $\mathcal{V}$ consists of a gauge field $\mathscr{A}_\mu$, a scalar field $\sigma$, a two-component Dirac spinor $\chi$, and another scalar field $D$,
all transforming in the adjoint representation of the gauge group $\mathcal{G}$. This is simply the dimensional reduction of the usual four-dimensional ${\cal N}=1$ vector multiplet. In particular, $\sigma$ arises from the zero mode of the component of the vector field in the direction along which one reduces. The matter fields $\Phi_a$ are chiral multiplets, consisting of a complex scalar $\phi_a$, a fermion $\psi_a$ and
an auxiliary scalar $F_a$, which may be in arbitrary representations $\mathcal{R}_a$ of $\mathcal{G}$.  An ${\cal N}=2$ Chern-Simons-matter Lagrangian
then consists of three terms:
\bea\label{action}
S \, =\, S_{\mathrm{CS}} + S_{\mathrm{matter}} + S_{\mathrm{potential}}~.
\eea

We will be interested in product gauge groups of the form $\mathcal{G} = \prod_{I=1}^\gauge U(N_I)$, 
where the Chern-Simons level for the $I$th factor $U(N_I)$ is $k_I\in \Z$. 
If $\mathcal{V}_I$ denotes the projection
of $\mathcal{V}$ onto the $I$th gauge group factor, then in component notation the Chern-Simons action, in Wess-Zumino gauge, takes the form
\bea
S_{\mathrm{CS}}\, = \, \sum_{I=1}^\gauge \frac{k_I}{4\pi}\int  \mathrm{Tr} \,\left( \mathscr{A}_I \wedge \diff \mathscr{A}_I + \frac{2}{3} \mathscr{A}_I\wedge \mathscr{A}_I\wedge \mathscr{A}_I - \bar\chi_I \chi_I +
2D_I\sigma_I \right)~,
\label{CSaction}
\eea
where the trace in (\ref{CSaction}) is normalized in the fundamental representation.

The matter kinetic term takes a simple form in superspace, namely
\bea
S_{\mathrm{matter}} & = & \int \diff^3 x\,  \diff^4\theta \sum_a \,  \mathrm{Tr} \,\bar \Phi_a \ex^{\mathcal{V}} \Phi_a\nonumber\\
&=& \int \diff^3 x \sum_a \mathscr{D}_\mu \bar\phi_a \mathscr{D}^\mu \phi_a - \bar\phi_a\sigma^2 \phi_a + \bar\phi_a D \phi_a + \mathrm{fermions}~. \label{mataction}
\eea
In the second line we have expanded into component fields, and we have
not written the terms involving the fermions $\psi_a$. $\mathscr{D}_\mu$ is the covariant derivative, and the auxiliary
fields $\sigma$ and $D$ are understood to act on $\phi_a$
in the appropriate representation $\mathcal{R}_a$. In this paper we shall mainly focus 
on theories with matter in bifundamental or adjoint representations of a gauge group 
$\mathcal{G} = \prod_{I=1}^\gauge U(N_I)$. In this case one can represent the gauge group and 
matter content by a quiver diagram with $\gauge$ nodes, with a directed arrow from node 
$I$ to node $J$ corresponding to a bifundamental field in the representation $\mathbf{N}_I\otimes 
\overline{\mathbf{N}}_J$; when $I=J$ this is understood to be the adjoint representation.

Finally, the superpotential term is
\bea
S_{\mathrm{potential}} & = &  \int \diff^3 x\,  \diff^2\theta  \, W(\Phi) + \mathrm{c}. \mathrm{c}.\nonumber\\
&=& -\int \diff^3 x \sum_a \, \left|\frac{\partial W}{\partial \phi_a}\right|^2 + \mathrm{fermions}~.
\label{potaction}
\eea
One typically takes the superpotential to be a polynomial in the chiral fields $\Phi_a$, and we have included the couplings (which in general will be renormalized) in the definition of $W$.

In \cite{Martelli:2008si,Hanany:2008cd} the moduli spaces of vacua of such theories were studied.  As usual this means that all scalar fields take 
constant vacuum expectation values, and one seeks absolute minima of the total potential. This is equivalent to imposing the 
F-terms $\diff W=0$, together with an appropriate K\"ahler quotient by the gauge group. The latter is more 
subtle than the corresponding quotient for four-dimensional $\mathcal{N}=1$ gauge theories, due to the effects of the Chern-Simons 
interactions.  The result is that for a large class of such theories, provided
\begin{equation}\label{CSsum}
\sum_{I=1}^\gauge k_I\, =\, 0~,
\end{equation}
and one takes the ranks $N_I=N$ to be equal\footnote{In fact this is not really necessary. For a detailed discussion in a particular class of examples, see \cite{Aharony:2008gk, 
Martelli:2009ga}.} for all $I=1,\ldots,\gauge$, then the moduli space of vacua contains the symmetric product $\mathrm{Sym}^N X$,
where $X$ is an affine Calabi-Yau four-fold variety. 

The simplest construction of $\mathcal{N}=2$ Chern-Simons-matter theories within
this class is to begin with the gauge group and matter content of a ``parent'' four-dimensional $\mathcal{N}=1$ gauge theory 
which has a Calabi-Yau \emph{three-fold} as Abelian moduli space \cite{Martelli:2008si}, and reinterpret this as a three-dimensional 
$\mathcal{N}=2$ theory by adding the Chern-Simons interactions (\ref{CSaction}). This has 
an elegant string theory interpretation \cite{Aganagic:2009zk}: if the initial four-dimensional gauge theory is the effective theory on $N$ D3-branes probing 
a Calabi-Yau three-fold singularity, then  one can T-dualize along a worldvolume direction to obtain 
a corresponding theory on D2-branes in Type IIA string theory. The addition of Chern-Simons couplings 
may then be understood in terms of turning on Ramond-Ramond two-form (and more generally four-form) flux 
and lifting to M-theory. One can also add D6-branes in this set-up, which introduces new chiral matter 
fields in fundamental representations of the gauge group factors 
\cite{Jafferis:2009th, Benini:2009qs}.
The resulting theories 
are generally conjectured to be low energy effective field theories on $N$ M2 branes probing a Calabi-Yau four-fold singularity $X$. 

In this gauge theory construction the Calabi-Yau four-fold $X$ is topologically a cone over a compact seven-manifold $Y$, and a Calabi-Yau metric on $X$ of the conical form
\bea\label{cone}
g_X & =&\diff r^2 + r^2 g_Y~,
\eea
implies that $Y$ is a Sasaki-Einstein manifold. The AdS/CFT correspondence conjectures that the IR limit of the 
Chern-Simons-matter theory, for fixed Chern-Simons levels and large $N$, is holographically dual to 
M-theory on the Freund-Rubin background AdS$_4\times Y$ with $N$ units of $\star G_4$-flux through $Y$, where 
$G_4$ is the M-theory four-form and $\star$ denotes the eleven-dimensional Hodge dual. 

We note that one can relax the condition (\ref{CSsum}), which leads to theories that are conjecturally dual to 
\emph{massive} Type IIA string theory backgrounds \cite{Gaiotto:2009mv}. This may be understood 
in the same context as \cite{Aganagic:2009zk}, via the effects of turning on Ramond-Ramond fluxes in Type IIA
on the Chern-Simons couplings on fractional branes. The results in this paper presumably 
extend to these gravity backgrounds also, although they are no longer described by Sasaki-Einstein geometry.

\subsection{Localization of the partition function to a matrix model}\label{localize}

In \cite{Kapustin:2009kz} it was shown that the partition function 
of a \emph{superconformal} Chern-Simons-matter theory on $S^3$ 
localizes to a matrix model. The derivation follows the usual method 
of localization: one adds an appropriately chosen $Q$-exact operator to the action of the theory, 
where $Q$ is a supercharge with $Q^2=0$. One can formally argue that this does not affect the partition function, but at the same time it localizes fields to certain constant values, for which the one-loop 
approximation is exact.
More precisely, in this case
the adjoint scalar $\sigma$ in the vector multiplet is localized to constant
field configurations, with all other fields being zero. 

For the 
theories described in section \ref{CSmatter}, the gauge group 
is of the product form $\mathcal{G}=\prod_{I=1}^\gauge U(N_I)$ 
and correspondingly $\sigma_I$ is a Hermitian $N_I\times N_I$ matrix. 
Up to gauge equivalence $\sigma_I$ is described by its $N_I$ real eigenvalues 
$\lambda^I_{i}$, where $i=1,\ldots,N_I$, and the matrix model 
in question is then a multi-matrix model for these eigenvalues. 
Such matrix models were subsequently studied by a number of authors \cite{Drukker:2010nc,   Suyama:2009pd,
Herzog:2010hf, Marino:2009jd, Suyama:2010hr}. In particular, the matrix model for the ABJM theory 
was solved in \cite{Marino:2009jd, Drukker:2010nc} in the limit in which 
$N$ is large and $k/N$ is held fixed. In \cite{Herzog:2010hf} 
instead the authors studied what we will refer to as the \emph{M-theory limit} 
of large $N$ and $k$ held fixed for the ABJM theory. In addition they studied 
some closely related theories with $\mathcal{N}=3$ superconformal symmetry, 
to which the same techniques apply. 

In the present paper we are interested in the more general situation in which 
one has some UV $\mathcal{N}=2$ Chern-Simons-matter theory, with action (\ref{action}), 
which one believes flows to a superconformal fixed point in the IR. In this case 
the results of \cite{Kapustin:2009kz} do not  directly apply, since at the fixed point the 
chiral matter fields $\Phi_a$ will in general not have canonical scaling dimensions 
$\Delta=\tfrac{1}{2}$. Fortunately, this problem has recently been addressed in 
\cite{Jafferis:2010un}. Here the author considered $\mathcal{N}=2$ Chern-Simons-matter 
theories with a choice of R-symmetry, with appropriate supersymmetry-preserving 
R-charge dependent couplings to the curvature of $S^3$,
and showed that this partition function 
on $S^3$ still localizes as in \cite{Kapustin:2009kz}. The resulting partition function 
then depends on this choice of R-symmetry. For a superconformal theory, one 
should of course choose this to be the superconformal R-symmetry of the IR theory.

It is straightforward to apply the localization results of \cite{Kapustin:2009kz, Jafferis:2010un} 
to the theories described in section \ref{CSmatter}. As already mentioned, 
the adjoint scalars $\sigma_I$ in the vector multiplet localize to constant field 
configurations, and using the gauge freedom we may parametrize these 
by their eigenvalues $\lambda^I_i$. All other fields, and in particular the 
chiral scalar fields $\phi_a$, are localized to zero. The partition function then localizes 
to the finite matrix integral
\bea\label{Z}
Z \, =\, \frac{1}{(\prod_{I=1}^\gauge N_I!)}\int \left( \prod_{I=1}^\gauge \prod_{i=1}^{N_I} 
\frac{\diff \lambda^I_i}{2\pi}\right) \exp\left[{\ii}\sum_{I=1}^\gauge \frac{k_I}{4\pi} \sum_{i=1}^{N_I}\left(\lambda^I_i\right)^2\right]
\exp\left[-F_{\mathrm{loop}}\right]~,
\eea
where the one-loop term is
\bea\label{Zloop}
\exp\left[-F_\mathrm{loop}\right] &=& \prod_{I=1}^\gauge \prod_{i\neq j} 2 \sinh \left(\frac{\lambda^I_i-\lambda^I_j}{2}\right)\cdot \exp[-F_{\mathrm{matter}}]~.
\eea
Here the first exponential term in (\ref{Z}) is simply that of the Euclideanized Chern-Simons action (\ref{CSaction}), localized 
onto the constant field configuration of the $\sigma_I$. The localization renders the one-loop approximation to the path integral 
exact, and the first term in (\ref{Zloop}) is precisely this one-loop contribution for the gauge sector of the theory. As usual, 
this is a regularized determinant, derived from the action (\ref{CSaction}). 

The matter sector 
contributes nothing at tree level, since the matter multiplets localize to zero, but there is a one-loop 
determinant factor given by \cite{Jafferis:2010un}
\bea\label{Zmatter}
\exp\left[-F_{\mathrm{matter}}\right] &=& \prod_{a}\, \mbox{det}_{\mathcal{R}_a} \exp\left[\ell\left(1-\Delta_a+\ii \sigma\right)\right]~.
\eea
Recall here that the index $a$ labels chiral matter fields $\Phi_a$ in the representation $\mathcal{R}_a$ of $\mathcal{G}$. 
We have denoted the conformal dimension/R-charge of $\Phi_a$ by $\Delta_a = \Delta[\Phi_a]$. 
In (\ref{Zmatter}), the determinant in the representation $\mathcal{R}_a$ is understood
 to be a product over weights $\rho$ in the weight-space decomposition of this representation, 
 and $\sigma$ is then understood to mean $\rho(\sigma)$. For example, for a bifundamental field 
$\Phi_a=\Phib$ in the representation $\mathbf{N}_I\otimes \overline{\mathbf{N}}_J$ we have
\bea
\exp\left[-F(\Phib)\right] &=& \prod_{i=1}^{N_I}\prod_{j=1}^{N_J} \exp\left[\ell\left(1-\Db+\ii \frac{\lambda^I_i-\lambda^J_j}{2\pi}\right)\right]~,
\label{FBI}
\eea
while for an adjoint field $\Phia$ for the gauge group factor $U(N_I)$ we have
\bea\label{Ad}
\exp\left[-F(\Phia)\right] &=& \prod_{i,j=1}^{N_I} \exp\left[\ell\left(1-\Da+\ii \frac{\lambda^I_i-\lambda^I_j}{2\pi}\right)\right]~.
 \eea
Finally, the function $\ell$ in (\ref{Zmatter}) arises from the (zeta function) regularized one-loop determinant of the 
matter action (\ref{mataction}), and is given explicitly by
\bea
\ell(z) & = &  -z \log\left(1-\ex^{2\pi\ii z}\right) + \frac{\ii}{2}\left[\pi z^2 + \frac{1}{\pi}\mathrm{Li}_2\left(\ex^{2\pi i z}\right)\right] - \frac{\ii\pi}{12}~.
\eea
Here $\mathrm{Li}_2(\zeta)$ denotes the dilogarithm function, defined as the analytic continuation of 
\bea
\mathrm{Li}_2(\zeta) &=& \sum_{m=1}^\infty \frac{\zeta^m}{m^2}~, \qquad |\zeta|<1~,
\eea
to $\C\setminus[1,\infty)$. 
The function $\ell(z)$  satisfies the simple differential equation
\bea
\frac{\diff \ell}{\diff z} &=& -\pi z \cot \pi z~,
\eea
together with the boundary condition that $\ell(0)=0$. For canonical scaling dimensions $\Delta=\tfrac{1}{2}$, this reduces 
to the original result of \cite{Kapustin:2009kz}.

\subsection{The M-theory limit of the free energy and volumes}

We are interested in computing the \emph{free energy}, $F=-\log Z$, for these matrix models in the limit in which
the Chern-Simons levels $k_I$, and in particular their greatest common divisor $k=\mathrm{gcd}\{k_I\}$, are held fixed and $N\rightarrow\infty$. This is the \emph{M-theory limit}, 
as opposed to the \emph{'t Hooft limit} in which $N/k$ is held fixed while $N\rightarrow\infty$. For simplicity, here we assume that 
the ranks of the gauge group factors are all equal, so $N_I=N$ for all $I=1,\ldots,\gauge$. 

Consider matrix models arising from Chern-Simons-matter theories which flow to superconformal fixed points, 
with holographic duals of the form AdS$_4\times Y$, as described at the end of  section \ref{CSmatter}. 
If the  $\Delta_a$ are taken to be the conformal dimensions/R-charges at this fixed point, then
the AdS/CFT correspondence predicts that to leading order as $N\rightarrow\infty$ \cite{Emparan:1999pm,Herzog:2010hf}
\bea\label{vols}
F &=& - \log Z \, = \,  N^{3/2}\sqrt{\frac{2\pi^6}{27\mathrm{Vol}(Y)}}~,
\eea
where $\mathrm{Vol}(Y)$ denotes the volume of the Sasaki-Einstein metric on $Y$. 
This formula follows from the saddle point approximation to the gravitational 
partition function on AdS$_4$, regularized using counterterm subtraction.
In particular, we see the famous
$N^{3/2}$ scaling of the free energy of $N$ M2 branes in the large $N$ limit in (\ref{vols}). 
This leads to a general concrete prediction: for a Chern-Simons-matter theory 
with Abelian moduli space $X$, which is a cone over $Y$,  the free energy of 
the corresponding localized matrix model is related to the volume of a Sasaki-Einstein metric on $Y$ via 
(\ref{vols}), in the M-theory large $N$ limit. 

This prediction has been confirmed for the ABJM theory in \cite{Marino:2009jd, Drukker:2010nc, 
Herzog:2010hf}, and in the latter reference even more remarkably it has been confirmed 
for a simple class of $\mathcal{N}=3$ Chern-Simons-matter theories \cite{Jafferis:2008qz} that are holographically 
dual to certain \emph{tri-Sasakian} seven-manifolds. The matrix models in these cases 
are somewhat simpler than the generic case described in the previous subsection, as originally 
noted in \cite{Kapustin:2009kz}. In particular, the combinations of $\ell$-functions 
in (\ref{Zmatter}) simplify to hyperbolic cosines for these $\mathcal{N}=3$ theories.  
Related to this, the fields all have canonical scaling dimensions of $\Delta=\tfrac{1}{2}$. 

In this paper we wish to extend the correspondence further by checking 
that (\ref{vols}) holds for $\mathcal{N}=2$ theories which are only conjectured to be superconformal 
in the IR limit, with non-canonical scaling dimensions $\Delta\neq \tfrac{1}{2}$. 
In turn, this acts as an AdS/CFT check on the field theory results of \cite{Jafferis:2010un}.
 In fact, we will provide non-trivial checks of our stronger conjecture (\ref{conjecture}).

\section{The M-theory limit of Chern-Simons-matter matrix models}\label{gensection}

In this section we discuss the M-theory large $N$ limit of the partition function (\ref{Z}). 
In sections \ref{Ansec} and \ref{SPPsec} we shall compute this explicitly for families of 
Chern-Simons-matter theories, and verify that (\ref{Z}) leads to 
precisely (\ref{vols}), where $\mathrm{Vol}(Y)$ is the volume of the 
appropriate Sasakian manifold.

\subsection{Massive adjoint fields}

Before beginning our general analysis, we pause to comment on how \emph{massive} adjoint fields 
should be treated in the partition function. As we shall see, this is relevant for the 
$\mathcal{A}_{n-1}$ theories discussed in section \ref{Ansec}. 
Clearly, a theory with massive adjoint scalar fields cannot itself be superconformal. 
However, after integrating these out the theory may flow to a superconformal 
fixed point. How are we to treat such fields in the UV partition function (\ref{Z})? 
 As we now explain, 
the matrix model partition function effectively integrates out these fields for us.

Consider an adjoint scalar field $\Phia$ for a gauge group $U(N)$. From (\ref{Ad}) this contributes
\bea\label{massAd}
\prod_{i,j=1}^{N} \exp\left[\ell\left(1-\Da+\ii \frac{\lambda_i-\lambda_j}{2\pi}\right)\right]
\eea
to the matrix model partition function. Since for a massive field $\Da=1$, this simplifies to
\bea
\exp[N\ell(0)]\cdot\prod_{i>j} \exp\left[\ell\left(\ii \frac{\lambda_{ij}}{2\pi}\right)+\ell\left(-\ii \frac{\lambda_{ij}}{2\pi}\right)\right]~,
\eea
where  we have defined $\lambda_{ij}=\lambda_i-\lambda_j$. However, we now recall 
that $\ell(0)=0$, and in fact it is easy to show more generally that
\bea
\ell(\ii u)\, +\, \ell(-\ii u) &=& 0~, \qquad u\in \R~.
\eea
Thus the contribution (\ref{massAd}) of a massive adjoint to the partition function is identically equal to 1. 

\subsection{The saddle point approximation}\label{saddle}

The strategy for evaluating (\ref{Z}) in the M-theory limit will be to use the saddle point, 
or stationary phase, approximation to the integral. Here $N^2$ plays the role 
of $1/\hbar$, so that the $N\rightarrow\infty$ limit is dominated by 
saddle point configurations that extremize the quantum effective action. 
This is a somewhat standard technique, and we refer the reader to the review 
\cite{Marino:2004eq} for further details.

The quantum effective action is 
\bea
F &= & F_{\mathrm{classical}} + F_{\mathrm{loop}}~,
\eea
where $F_{\mathrm{loop}}$ is defined by (\ref{Zloop}), (\ref{Zmatter}) and
\bea
F_{\mathrm{classical}} &=& -\ii\sum_{I=1}^\gauge \frac{k_I}{4\pi} \sum_{i=1}^{N_I}\left(\lambda^I_i\right)^2
\eea
is the localized Chern-Simons action. The saddle point approximation to the partition function (\ref{Z}) is 
dominated by solutions to the equations of motion $\partial F/\partial\lambda_i^{I}=0$.
At this point it will be convenient to assume that the matter content is described by 
a quiver diagram with $N_I=N$, $I=1,\ldots, \gauge$, so that in particular all matter fields are in bifundamental or 
adjoint representations of the gauge group $U(N)^\gauge$.
 A straightforward computation then gives
\bea\label{genEOM}
-\frac{\partial F}{\partial\lambda_i^{I}} &=& \frac{\ii k_I}{2\pi} \lambda_i^{I} + \sum_{j\neq i}\coth 
\tfrac{\lambda_i^{I} - \lambda^I_j}{2}\\
&&- \frac{\ii}{2}\sumItoJ \sum_{j=1}^N \left(1-\Delta_{IJ} + \ii \tfrac{\lambda^I_i-\lambda^J_j}{2\pi}\right) \cot \left(\pi(1-\Delta_{IJ}) + \ii \tfrac{\lambda^I_i-\lambda^J_j}{2}\right)\nn\\
&& +\frac{\ii}{2}\sumJtoI \sum_{j=1}^N \left(1-\Delta_{JI} - \ii \tfrac{\lambda^I_i-\lambda^J_j}{2\pi}\right) \cot \left(\pi(1-\Delta_{JI}) - \ii \tfrac{\lambda^I_i-\lambda^J_j}{2}\right)~.\nn
\eea
Here the index $I$ is fixed, and the sums are over outgoing arrows $a=(I\rightarrow J)$ or incoming arrows 
$a=(I\leftarrow J)$. The conformal dimension of a bifundamental field from node $I$ to node $J$ is denoted
$\Delta_a=\Delta_{IJ}$. 
Notice that an adjoint field with $I=J$ contributes to both terms in (\ref{genEOM}).

Let us analyze these equations of motion assuming that 
the eigenvalues grow in the large $N$ limit. More precisely, as we shall see the saddle point 
approximation is in fact dominated by a \emph{complex} solution, in which the 
eigenvalues $\lambda_i^{I}$ have an imaginary part that is small compared to 
the real part. This is a common phenomenon: one is effectively deforming the real integral 
in (\ref{Z}) into the complex plane in order to find the steepest descent. Then more precisely 
we are assuming that the real parts $\xi_i^{I}=\Re \lambda_i^{I}$ of the eigenvalues grow\footnote{We shall be more 
precise about this later.} with $N$, 
where we introduce the real and imaginary parts:
\bea
\lambda^I_i &=& \xi^{I}_i + \ii  y^{I}_i~.
\eea
Using
\bea
\cot (a+\ii b )&=& \frac{1-\ii \tan a \tanh b}{\tan a + \ii \tanh b}~,
\eea
together with 
\bea
\tanh w & = &  \sgn(\Re w) + \mathcal{O}\left(\ex^{-2|\Re w|}\right)~,
\eea
we see that we may approximate $\cot (a+\ii b)\simeq -\ii \, \sgn (\Re b)$ for $|\Re b|\gg 0$. Using this,  the equation of motion (\ref{genEOM}) simplifies to
\bea
-\frac{\partial F}{\partial\lambda_i^{I}} &\simeq & \frac{\ii k_I}{2\pi} \lambda_i^{I} +\sum_{j\neq i}\sgn\left(
\xi_i^{I} - \xi^I_j\right)\nn \\
&&- \frac{1}{2}\sumItoJ \sum_{j=1}^N \left(1-\Delta_{IJ} + \ii \tfrac{\lambda^I_i-\lambda^J_j}{2\pi}\right)\sgn\left(\xi^I_i-\xi^J_j\right)\nn\\
&&-\frac{1}{2}\sumJtoI \sum_{j=1}^N \left(1-\Delta_{JI} - \ii \tfrac{\lambda^I_i-\lambda^J_j}{2\pi}\right)\sgn \left(\xi^I_i-\xi^J_j\right)~.\label{approxEOM}
\eea

We now introduce the \emph{continuum limit}. Here the sums over $N$ eigenvalues tend to Riemann integrals 
as $N\rightarrow \infty$. Again, this procedure is somewhat standard, and a review may be found in 
\cite{Marino:2004eq}. We begin by defining
functions $\xi^I(s)$, $y^I(s)$, so $\xi^I,y^I:[0,1]\rightarrow \R$, via
\begin{equation}\label{con}
\xi^I\left[\tfrac{1}{N}\left(i-\tfrac{1}{2}\right) \right]\, = \, \xi_i^{I}~,\qquad y^I\left[\tfrac{1}{N}\left(i-\tfrac{1}{2}\right)\right]\,  =\,  y_i^{I}~, \qquad i=1,\ldots, N~.
\end{equation}
In the continuum limit where $N\rightarrow\infty$ the sums over $N$ become Riemann integrals, so for example
\begin{equation}
\frac{1}{N}\sum_{j=1}^N f(\xi_j) \longrightarrow \int_0^1 f(\xi(s))\, \diff s~.
\end{equation}
For each $I=1,\ldots, \gauge$ we may also introduce the \emph{density} $\rho_I(\xi)$ via
\bea
\rho_I(\xi^I(s))\diff \xi^I &=& \diff s~.
\eea
These are $\gauge$ functions of a single real variable. Assuming that 
the real parts of the eigenvalues dominate over the imaginary parts at large $N$, the leading order 
term in  (\ref{approxEOM}) in the continuum limit gives
\bea\label{leadingEOM}
\sumItoJ \int \left(\xi-\xi'\right) \rho_J(\xi') \sgn (\xi-\xi')\diff \xi' \,  = \!\!\!\! \sumJtoI\int \left(\xi-\xi'\right) \rho_J(\xi') \sgn(\xi-\xi')\diff \xi' ~
\eea
Differentiating this twice with respect to $\xi$ then implies
\bea\label{rhos}
\sumItoJ \rho_J(\xi)& =& \sumJtoI \rho_{J}(\xi)~, \qquad I=1,\ldots,\gauge~.
\eea
Remarkably, these $\gauge$ linear equations for the $\gauge$ density functions $\rho_I(\xi)$ 
take the form of ABJ anomaly conditions, when thought of in the four-dimensional context. 
If one considers the parent $\mathcal{N}=1$ four-dimensional quiver gauge theory, then 
it is a general result that there is a $(b_3(Y_5)+1)$-dimensional space of solutions to (\ref{rhos}); that is,
 the skew part of the adjacency matrix of the quiver has a kernel of dimension $b_3(Y_5)+1$.
Here $b_3(Y_5)$ is the third Betti number of $Y_5$, which is the link of the Calabi-Yau \emph{three-fold} 
singularity of the parent theory vacuum moduli space. Given (\ref{rhos}), the leading 
order part of the equation of motion (\ref{leadingEOM}) is then zero.

The next-to-leading order term in (\ref{genEOM}) then gives
\bea
0 & = & \int \rho_I(\xi')\, \sgn(\xi-\xi')\, \diff\xi' \nn\\
&& - \frac{1}{2}\sumItoJ\int 
\left(1-\Delta_{IJ} - \tfrac{y^I(\xi) - y^J(\xi')}{2\pi}\right)\rho_J(\xi')\, \sgn (\xi-\xi')\, \diff \xi'\nn\\
&& - \frac{1}{2}\sumJtoI\int 
\left(1-\Delta_{JI} +\tfrac{y^I(\xi) - y^J(\xi')}{2\pi}\right)\rho_J(\xi')\, \sgn (\xi-\xi')\, \diff \xi'~.
\eea
Using (\ref{rhos}) the $y^I(\xi)$ terms cancel, and differentiating gives
\bea
2\rho_I(\xi) \, =\,  \sumItoJ\left(1-\Delta_{IJ} +\tfrac{y^J(\xi)}{2\pi}\right)\rho_J(\xi)
+  \sumJtoI\left(1-\Delta_{JI} -\tfrac{y^J(\xi)}{2\pi}\right)\rho_J(\xi)~.
\eea

At this point we shall make the assumption that
\bea\label{assume}
\rho_I(\xi) &=& \rho(\xi)
\eea
holds for all $I=1,\ldots,\gauge$. This condition is satisfied for all the $\mathcal{N}\geq 3$ examples discussed in \cite{Herzog:2010hf}, 
and as we shall see later also holds for the $\mathcal{N}=2$ $\mathcal{A}_{n-1}$ theories  by a symmetry argument. 
As we shall also see, without the constraint (\ref{assume}) the behaviour of the matrix model is qualitatively different. 
Notice that for theories with parents for which $b_3(Y_5)=0$, the condition (\ref{assume}) \emph{necessarily} holds 
as the skew adjacency matrix has a one-dimensional kernel. This kernel is due to  gauge anomaly cancellation 
in four dimensions, which for equal ranks $N_I=N$ is equivalent to the number of incoming arrows 
equalling the number of outgoing arrows at each node. With  (\ref{assume}), 
the last equation becomes 
\bea\label{NS}
2 &=& \sumItoJ\left(1-\Delta_{IJ} +\tfrac{y^J(\xi)}{2\pi}\right) +  \sumJtoI\left(1-\Delta_{JI} -\tfrac{y^J(\xi)}{2\pi}\right)~.
\eea
One can now sum this equation over $I$. This gives
\bea\label{summy}
2G& =& 2 \sumarrows (1-\Delta_{IJ})+\sumarrows\frac{y^J(\xi)-y^I(\xi)}{2\pi}~.
\eea
Here the sums are over \emph{all} matter content, or equivalently arrows $a=(I\rightarrow J)$ in the quiver. The first factor of 2 arises because we double count every arrow in the quiver (each arrow is incoming and outgoing to precisely 
one node each). The last term in (\ref{summy}) is then zero due to four-dimensional gauge anomaly cancellation of the parent theory:  note 
that for a fixed node $I$ an outgoing arrow contributes $-y^I(\xi)$, while an incoming arrow contributes 
$+y^I(\xi)$. We thus derive the following constraint on the conformal dimensions:
\bea
G &=& \sumarrows (1-\Delta_{IJ})~.
\eea

Given (\ref{assume}), it follows that $\Re \lambda_i^I=\Re \lambda_i^J\equiv \xi_i$ holds for all $I$ and $J$. We may thus 
write
\bea
\lambda_i^I &=& \xi_i + \ii y^I_i~,
\eea
with $\xi_i$, $y_i^I$ both real. At this point it is also convenient to order the eigenvalues in such a way that $\xi_i$ is monotonically 
increasing with $i$. We may then simplify the expression (\ref{FBI}) for $F(\Phib)$ for a generic bifundamental 
field $\Phib$. 
We first rearrange the terms in the products, 
separating the terms with $i>j$ and $i<j$ from those with $i=j$:
\begin{equation}
 F(\Phib) \ = \ F_1(\Phib) + F_2(\Phib)~.
\end{equation}
Here 
\bea\label{F1}
F_1(\Phib) &=& \sum_{i=1}^N \ell\left(1-\Db-\frac{y^I_{i}-y^J_i}{2\pi}\right)~.
\eea
In fact this term will be subleading in the large $N$ expansion, as we shall see momentarily. Using 
the expressions for $\ell_\pm(z)$ in Appendix \ref{expansions}, we also compute
\bea\label{F2bi}
F_2 (\Phib)&= &\frac{1}{2}\sum_{i\neq j} \mathrm{sgn}(\xi_i -\xi_j)
\left( 1-\Delta^{I,J} +\ii \frac{\lambda^I_i-\lambda^J_j}{4\pi}\right) (\lambda_i^I - \lambda_j^J) + \nn\\
&&\mbox{sums of exponentials}~.
\eea
Here the sums of exponentials are precisely  the sums over $m$ in equations 
(\ref{ellpm}). 
Of course, differentiating (\ref{F2bi}) with respect to $\lambda^I_i$ leads to the corresponding term in the 
leading order equation of motion (\ref{approxEOM}).

\subsection{Quantum effective action: non-chiral theories}

For a general Chern-Simons-matter theory, the terms written in (\ref{F2bi}) are at leading order 
in the large $N$ expansion, and generically do not cancel on summing over all matter content. 
In contrast to this, for the $\mathcal{N}\geq 3$ examples studied in \cite{Herzog:2010hf} 
these terms precisely cancelled. More generally, it is straightforward to see that this 
cancellation always occurs for a \emph{non-chiral} Chern-Simons-matter theory. 
In terms of quivers, this means that for every arrow from node $I$ to node $J$, there is a 
corresponding arrow from node $J$ to node $I$. 
At this point it is expedient to therefore restrict to non-chiral theories. The matrix models in question 
will then closely resemble that for the ABJM theory, but will nevertheless still be general enough 
to allow for non-trivial $\mathcal{N}=2$ theories with anomalous (and in fact irrational) dimensions $\Delta_a$ 
of the chiral matter fields. 

Requiring that the Chern-Simons-matter theory be non-chiral has a number of interesting consequences. 
Firstly, the terms depending on $y^J_i$ in (\ref{NS}) cancel, leaving us with the following set of constraints 
on conformal dimensions:
\bea\label{NSVZ}
2 &=& 2\sumIfromtoJ\left(1-\Delta_{IJ}\right) ~,
\eea
where  we have assumed\footnote{For the explicit examples that we shall study in this paper this will 
follow from symmetry; more generally one might have to relax this condition.} that $\Delta_{IJ}=\Delta_{JI}$ for each bifundamental pair.
These are precisely the conditions imposed by setting the $\gauge$ NSVZ beta functions in the four-dimensional 
parent theory to zero. Remarkably, the same conditions must hold for non-chiral Chern-Simons-matter theories, 
at least under the assumptions we have made thus far. 
It is then straightforward to check that the terms written in (\ref{F2bi}) cancel, leaving only the exponential 
sums. More precisely, the quadratic and constant terms cancel pairwise between the two arrows going between $I\leftrightarrow J$, 
while the linear terms only cancel on summing over the whole quiver, where one must also include the contribution 
from the gauge sector one-loop term in (\ref{Zloop}).

Thus the terms written explicitly in (\ref{F2bi}) cancel for a non-chiral quiver, and we are left with the 
sums of exponentials. For a fixed pair of arrows going between nodes $I\leftrightarrow J$, we compute the contribution
\bea\label{F2}
F_2(\Phi_{I\leftrightarrow J})&=& \sum_{i>j} \sum_{m=1}^\infty \ex^{-m\xi_{ij}}\Bigg\{
\frac{1}{\pi m}\Big[\xi_{ij} + \frac{1}{m}\Big]\sin 2\pi m(1-\Delta)\left[\ex^{\ii m (y^I_j-{y}^J_i)}+\ex^{-\ii m (y^I_i-{y}^J_j)}\right]
\nn\\
&&-\frac{2}{m}(1-\Delta)\cos 2\pi m(1-\Delta)\left[\ex^{\ii m (y^I_j-{y}^J_i)}+\ex^{-\ii m (y^I_i-{y}^J_j)}\right]\\
&& +\frac{\ii}{\pi m} \sin 2\pi m(1-\Delta)\left[(y_i^I-y^J_j)\ex^{-\ii m (y^I_i-{y}^J_j)} - 
(y^I_j-y^J_i)\ex^{\ii m (y^I_j-{y}^J_i)}\right] \Bigg\}~.\nn
\eea
Here $\Delta\equiv \Delta_{IJ}=\Delta_{JI}$, and we have defined 
\bea
\xi_{ij} &=& \xi_i - \xi_j~.
\eea

In the continuum limit of the previous subsection, we have $y^I=y^I(\xi)$ with density function 
\begin{equation}
\rho(\xi)\diff \xi = \diff s~.
\end{equation}
It is straightforward to apply this to the classical part of the action:
\begin{eqnarray}\label{classxi}
F_{\mathrm{classical}} &=& -{\ii }\sum_{I=1}^\gauge \frac{k_I}{4\pi}\sum_{i=1}^N \left(\lambda^I_i\right)^2 \, = \, 
\frac{1}{2\pi}\sum_{I=1}^\gauge \sum_{i=1}^N k_I \xi_i y^I_i + \mbox{lower order}\nn\\
&\longrightarrow& \frac{N}{2\pi}\int \xi \rho(\xi) \sum_{I=1}^G k_Iy^I(\xi)\, \diff \xi~.
\end{eqnarray}
At this point we will be more precise about the growth of $\xi_i$. Following \cite{Herzog:2010hf}, we write the ansatz
\bea
\xi_i &=& N^\alpha x_i~
\eea
for the leading order behaviour of the real parts of the eigenvalues, 
where $\alpha>0$ is some real constant. Then more precisely the last equation 
(\ref{classxi}) becomes
\bea\label{Fcl}
F_{\mathrm{classical}} &=& \frac{N^{1+\alpha}}{2\pi}\int_{x_1}^{x_2} x \rho(x) \sum_{I=1}^G k_I y^I(x)\, \diff x+ {o}(N^{1+\alpha})~.
\eea
We then notice that (\ref{F1}) is $\mathcal{O}(N)$, and is thus \emph{subleading} to the classical 
action, assuming $\alpha>0$. Of course, this is intuitively clear, since this term came from 
$i=j$ in the original sum over both $i$ and $j$, and thus should be measure zero 
compared to this latter term, in the continuum limit. 

We turn next to the leading order contribution to (\ref{F2}) in the continuum limit. In order to evaluate this, notice that since the sum over 
$i>j$ leads to a double integral with $x-x'>0$, 
the latter may be evaluated in  the large $N$ limit essentially using the representation of the delta function 
\bea
\delta (x) & =& \lim_{c\to \infty} \frac{c}{2} \ex^{-c|x|}~.
\eea
More precisely, consider the general equality
\begin{eqnarray}\label{identity}
\int_{x_1}^ x \diff x' \ex^{-mN^\alpha (x-x')} f(x,x') &=&  \frac{1}{mN^\alpha}\left[\ex^{-mN^\alpha (x-x')} 
f(x,x')\right]_{x_1}^x \nn \\
&&- \frac{1}{mN^\alpha} \int_{x_1}^x \diff x' \ex^{-mN^\alpha (x-x')} \frac{\diff}{\diff x'}f(x,x')~,
\end{eqnarray}
where we have made a trivial integration by parts. The first term is simply $\frac{1}{mN^{\alpha}}f(x,x)$ 
plus a term which is \emph{exponentially} suppressed in the large $N$ limit. One has to use this 
identity twice on the first term of (\ref{F2}), proportional to $\xi_{ij}=N^\alpha x_{ij}$, and once 
for each of the remaining terms, to derive the leading order result
\bea\label{F2fourier}
F_2(\Phi_{I\leftrightarrow J}) &=& \frac{4 N^{2-\alpha}}{\pi}\int_{x_1}^{x_2} (\rho(x))^2\diff x \sum_{m=1}^\infty 
\Bigg[\frac{1}{m^3}\sin 2\pi m(1-\Delta)\cos m\left[y^I(x)-y^J(x)\right]\nn\\
&&-\frac{\pi}{m^2}(1-\Delta)\cos 2\pi m(1-\Delta)\cos m\left[y^I(x)-y^J(x)\right]\nn\\
&&+ \frac{1}{2m^2}\left[y^I(x)-y^J(x)\right]\sin 2\pi m(1-\Delta)\sin m\left[y^I(x)-y^J(x)\right]\Bigg]~.
\eea
The crucial point here is that the first term in the first line of (\ref{F2})
is naively of order $N^2$. However, the identity (\ref{identity}) implies that this leading 
contribution is itself identically zero. One should then worry about subleading corrections to this term
coming from approximating the sum over $N$ eigenvalues with an integral. However, one can 
use the general estimate
\bea
\left|\int_0^1 f(x(s))\diff s - \frac{1}{N}\sum_{i=1}^N f(x_i)\right| & \leq & \frac{\sup_{s\in[0,1]}|f''(x(s))|}{24 N^2}~,
\eea
with $x_i$ related to $x(s)$ as in (\ref{con}),
to show this correction is subleading to (\ref{F2fourier}).

It is straightforward to similarly compute the leading order contribution to the partition function
for an \emph{adjoint} field of conformal dimension $\Delta$:
\bea\label{F2adjoint}
F_2(\Phia) &=& \frac{2N^{2-\alpha}}{\pi}\int_{x_1}^{x_2}(\rho(x))^2\diff x\sum_{m=1}^\infty
\Bigg[\frac{1}{m^3}\sin 2\pi m(1-\Delta)\nn \\
&&-\frac{\pi}{m^2}(1-\Delta)\cos 2\pi m (1-\Delta)\Bigg]~.
\eea
The leading order contribution for the gauge sector one-loop term in (\ref{Zloop}) is
\bea\label{F2gauge}
F_2(\mbox{gauge})&=& 2GN^{2-\alpha}\int_{x_1}^{x_2}(\rho(x))^2\diff x\sum_{m=1}^\infty \frac{1}{m^2}\nn\\
&=& \frac{\pi^2 GN^{2-\alpha}}{3}\int_{x_1}^{x_2}(\rho(x))^2\diff x~.
\eea
Here we have used the expansion (\ref{wexp}); recall that the linear term in (\ref{wexp}) has already been 
cancelled (see the paragraph after equation (\ref{NSVZ})). Notice that each $U(N)$ gauge group factor 
makes the same contribution, hence the overall factor of $G$ in (\ref{F2gauge}).

The sums over $m$ in (\ref{F2fourier}) and (\ref{F2adjoint})  may be evaluated in closed form using simple Fourier series results. 
We shall do this for our examples in the following sections.

\section{The $U(N)^2$ $\mathcal{A}_{n-1}$ theories}
\label{Ansec}

In this section we study the M-theory limit of the partition function for a particular class of 
$\mathcal{N}=2$ non-chiral Chern-Simons-matter theories.

\subsection{The quiver theories}
\label{An}

\begin{figure}[ht!]
\epsfxsize = 4.5cm
\centerline{\epsfbox{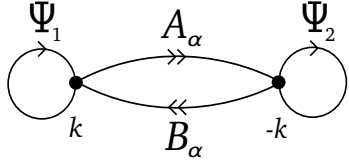}}
\caption{The $\mathcal{A}_{n-1}$ Chern-Simons quiver.}
\label{figv}
\end{figure}

In \cite{Martelli:2009ga} a particular family of Chern-Simons-matter theories was studied in 
considerable detail. For these theories the gauge group is $\mathcal{G}=U(N)_k\times U(N)_{-k}$, 
so that the number of gauge group factors is $\gauge=2$, and we have denoted the Chern-Simons levels as $k_1=k=-k_2$. 
The matter content consists of bifundamental fields $A_\alpha$, $B_\alpha$, $\alpha=1,2$, 
transforming in the ${\mathbf{N}}\otimes \overline{\mathbf{N}}$  and $\overline{\mathbf{N}}\otimes {\mathbf{N}}$ 
representations of the two gauge group factors, respectively, together with 
adjoint fields $\Psi_1$, $\Psi_2$ for each. The superpotential is
\bea\label{superpotential}
W \, = \, \mathrm{Tr}\, \left[s \left((-1)^{n}\Psi_2^{n+1}+\Psi_1^{n+1}\right) + \Psi_1(A_1B_1+A_2B_2)+
\Psi_2(B_1A_1+B_2A_2)\right]~,
\eea
where $s$ is a coupling constant and $n\in\mathbb{N}$ is a positive integer.  The superpotential is invariant under 
an $SU(2)$ flavour symmetry under which the adjoints $\Psi_I$ 
are singlets and both pairs of  bifundamentals $A_\alpha, B_\alpha$ transform as doublets. 
There is also a $ \zflip $ symmetry which 
exchanges $\Psi_1\leftrightarrow \Psi_2$, $A_\alpha\leftrightarrow B_\alpha$, 
$s\leftrightarrow (-1)^n s$. The quiver diagram is shown in Figure \ref{figv}.

The case $n=1$ is special, since then the first 
two terms in (\ref{superpotential}) give a mass to the adjoint fields $\Psi_1$, $\Psi_2$. 
At energy scales below this mass, we may therefore integrate out these fields.
On setting  $s=k/8\pi$, one recovers the ABJM theory \cite{ABJM} with quartic superpotential
\bea
W_{\mathrm{ABJM}} &= & \frac{4\pi}{k}(A_1B_2A_2B_1 - A_1B_1A_2B_2)~.
\eea
This theory is superconformal with enhanced $\mathcal{N}=6$ supersymmetry. 
We may thus regard this family of theories, which we refer to as the 
$\mathcal{A}_{n-1}$ theories, as a generalization of the ABJM theory.  

The Abelian ($N=1$) moduli space of vacua of this theory, for $k=1$, 
is the hypersurface singularity \cite{Martelli:2009ga}
\bea\label{Xn}
X_n &=& \left\{z_0^n + z_1^2+z_2^2+z_3^2+z_4^2=0\right\}\subset \C^5~.
\eea
For $n=1$ one easily sees that $X_1\cong \C^4$, as one expects since 
the Abelian ABJM theory with $k=1$ describes a single M2 brane in flat spacetime. 
For $n>1$ there is an isolated singularity of $X_n$ at the origin in $\C^5$. 
For $k>1$ one obtains instead the quotient $X_n/\Z_k$, where the $\Z_k$ action is 
free away from the origin. Then $X_n/\Z_k$ is a cone over a smooth compact Sasakian 
seven-manifold $Y_n/\Z_k$. Recall that a Sasakian metric on an odd-dimensional manifold 
$Y$ is equivalent to the cone metric (\ref{cone}) being K\"ahler.

The $\mathcal{A}_{n-1}$ theories are not classically conformally invariant, as one 
immediately sees from the non-quartic superpotential (\ref{superpotential}). 
However, one might be tempted to conjecture that the quantum theories 
flow to a strongly coupled interacting superconformal fixed point in the IR, 
at which (\ref{superpotential}) is marginal. At such a superconformal fixed 
point the theory develops an R-symmetry, where the superpotential necessarily 
has R-charge 2. Since a chiral superfield $\Phi_a$ saturates the BPS bound 
that $R[\Phi_a]=\Delta[\Phi_a]$, where $\Delta[\Phi_a]$ denotes the conformal dimension, 
it follows that at such a superconformal fixed point we must have
\bea\label{confdims}
&& \Db \, \equiv \, \Delta[A_\alpha] \, = \, \Delta[B_\alpha] \, = \, \frac{n}{n+1}~,\nn\\
&& \Da \, \equiv \, \Delta[\Psi_1] \, = \, \Delta[\Psi_2]\, = \, \frac{2}{n+1}~.
\eea
As explained in \cite{Martelli:2009ga}, this is problematic for $n>2$, since 
$\mathrm{Tr}\, \Psi_I$ would  then be gauge-invariant chiral primary 
operators with conformal dimensions $\Delta[\mathrm{Tr}\, \Psi_I]=\tfrac{2}{n+1}$, 
and this violates the \emph{unitarity bound} of $\Delta\geq \tfrac{1}{2}$, 
with equality only for a \emph{free} field, for $n>2$. Thus only for $n=2$ 
could this conjecture be true. In \cite{Martelli:2009ga} we conjectured instead 
that the $\mathcal{A}_{n-1}$ theories all flow to the \emph{same} conformal fixed 
point for $n>2$, in which the coupling constant $s=0$ and thus the 
$\Psi_I^{n+1}$ terms are absent in the superpotential (\ref{superpotential}).
In this case the vacuum moduli space of the theory is 
$\C\times \mathrm{Con}/\Z_k$, where $\mathrm{Con}=\{xy=uv\}\subset\C^4$ 
is the usual conifold three-fold singularity \cite{Martelli:2009ga}, and
the link $Y$ certainly admits a (singular) Sasaki-Einstein metric. 
We shall present further field theory evidence for this conjecture in section \ref{nmorethantwo}.

There is a gravitational dual to this \cite{Gauntlett:2006vf}. 
In general the existence of Sasaki-Einstein metrics, for example on 
links of hypersurface singularities, is a difficult unsolved problem. 
In \cite{Gauntlett:2006vf} it was pointed out that there are some simple 
holomorphic obstructions, which moreover have AdS/CFT dual interpretations. 
In particular, the unitarity bound obstruction above is dual to the \emph{Lichnerowicz} 
obstruction to the existence of Sasaki-Einstein metrics described in \cite{Gauntlett:2006vf}. 
 One can then show that no Sasaki-Einstein metric exists on the link $Y_n$ of 
$X_n$ in (\ref{Xn}), for $n>2$. On the other hand, for $n=2$ the quadric hypersurface 
certainly admits a Ricci-flat K\"ahler cone metric, where the Sasaki-Einstein metric on 
$Y_2$ is the homogeneous $V_{5,2}=SO(5)/SO(3)$ metric. In spite of this, 
certainly there exist \emph{Sasakian}, but \emph{non}-Einstein, metrics on 
$Y_n$, and the volumes of these manifolds are then independent of the choice of such a metric 
\cite{Martelli:2006yb}. One easily calculates this volume using the techniques in the latter reference, 
to obtain
\bea\label{Voln}
\mathrm{Vol}(Y_n) &=& \frac{(n+1)^4}{16n^3}\mathrm{Vol}(S^7) \, = \,  \frac{(n+1)^4\pi^4}{48n^3}~.
\eea
We stress again that for $n=2$ this is the volume of the homogeneous Sasaki-Einstein manifold 
$V_{5,2}$, while for $n>2$ it is the volume of \emph{any} Sasakian metric on
the link of $X_n$ (with canonical choice of Reeb vector field), although there is no 
Sasaki-Einstein metric.

\subsection{The partition function}

The matrix model for the $\mathcal{A}_{n-1}$ theories is easily obtained 
from the general formula (\ref{Z}). We have gauge group $\mathcal{G}=U(N)_k\times U(N)_{-k}$,
so that $G=2$, and the partition function is
\bea\label{Zn}
Z[\mathcal{A}_{n-1}] \,  = \, \frac{1}{(N!)^2}\int \left( \prod_{i=1}^N
\frac{\diff \lambda_i}{2\pi}\frac{\diff \tilde\lambda_i}{2\pi}\right) \exp\left[\frac{\ii k}{4\pi} \sum_{i=1}^{N}\left(\lambda^2_i - \tilde\lambda^2_i\right)\right]
\exp\left[-F_{\mathrm{loop}}\right]~,\eea
where
\bea\label{Zloopn}
&&\exp\left[-F_{\mathrm{loop}}\right] \, = \,  \prod_{i\neq j} 2 \sinh \frac{\lambda_{ij}}{2}\cdot  2\sinh \frac{\tilde\lambda_{ij}}{2}\cdot 
\prod_{i,j}\exp\Bigg[2\ell\left(1-\Db+\ii\frac{\hat\lambda_{ij}}{2\pi}\right) \nn\\
&&+ 2\ell\left(1-\Db-\ii\frac{\hat\lambda_{ij}}{2\pi}\right)+\ell\left(1-\Da+\ii\frac{\lambda_{ij}}{2\pi}\right)+\ell\left(1-\Da+\ii\frac{\tilde\lambda_{ij}}{2\pi}\right)\Bigg]~.
\eea
Here we have denoted the two sets of eigenvalues as $\lambda^{1}_i=\lambda_i$, $\lambda^{2}_i=\tilde\lambda_i$, and have introduced the 
notations
\bea\label{rels}
\lambda_{ij} \, =\,  \lambda_i - \lambda_j~, \qquad \tilde\lambda_{ij}\, =\, \tilde\lambda_i - \tilde\lambda_j~, \qquad \hat\lambda_{ij}\,  =\,  \lambda_i - \tilde\lambda_j~.
\eea
We also note from (\ref{confdims}) that, provided $s \neq 0$, the marginality of the superpotential (\ref{superpotential}) implies
\bea\label{cds}
1-\Db \, =\, \frac{1}{n+1}~, \qquad 1-\Da \, = \, \frac{n-1}{n+1}~.
\eea
For the time being we shall assume that $s\neq 0$, and thus that (\ref{cds}) hold, and return to the case that $s=0$ 
in section \ref{nmorethantwo}.
The two factors of 2 before the $\ell$ functions in (\ref{Zloopn}) arise from the sum over $\alpha=1,2$ on the bifundamental fields $A_\alpha$, $B_\alpha$.

\subsection{Symmetries and saddle point equations}

For the $\mathcal{A}_{n-1}$ theories the equation of motion (\ref{genEOM}) may be written as
\begin{eqnarray}\label{EOMn}
-\frac{\partial F}{\partial\lambda_i} &=& \frac{\ii k}{2\pi} \lambda_i + \sum_{j\neq i} \coth \frac{\lambda_{ij}}{2} 
+  \sum_{j=1}^N \frac{ \tfrac{\hat\lambda_{ij}}{2\pi}\sin \tfrac{2\pi}{n+1} - \tfrac{2}{n+1} \sinh \tfrac{\hat\lambda_{ij}}{2} \cosh \tfrac{\hat\lambda_{ij}}{2}}{\sin^2 \tfrac{\pi}{n+1}+\sinh^2 
\tfrac{\hat\lambda_{ij}}{2}}\nn\\
&& +\frac{1}{2} \sum_{j=1}^N \frac{ \tfrac{\lambda_{ij}}{2\pi}\sin \tfrac{2\pi (n-1)}{n+1} - \tfrac{2(n-1)}{n+1} \sinh \tfrac{\lambda_{ij}}{2} \cosh \tfrac{\lambda_{ij}}{2}}{\sin^2 \tfrac{\pi(n-1)}{n+1}+\sinh^2 
\tfrac{\lambda_{ij}}{2}}~.
\end{eqnarray}
The corresponding equation of motion for $\lambda^2_i=\tilde\lambda_i$ is obtained via the replacements $\lambda_i\leftrightarrow\tilde\lambda_i$, $k\rightarrow -k$. This is a remnant of 
the  $ \zflip $ symmetry which 
exchanges $\Psi_1\leftrightarrow \Psi_2$ and $A_\alpha\leftrightarrow B_\alpha$. 
In fact we note the following symmetries:
\begin{enumerate}
\item The equations of motion for $\lambda_i$ and $\tilde\lambda_i$ are interchanged via $\lambda_i\leftrightarrow \tilde\lambda_i$, $k\leftrightarrow -k$, which 
as mentioned is the $\Z_2^{\mathrm{flip}}$ symmetry of the $\mathcal{A}_{n-1}$ Chern-Simons-matter theory.
\item The equations of motion for $\lambda_i$ and $\tilde\lambda_i$ are interchanged via $\tilde\lambda_i\leftrightarrow \bar{\lambda}_i$. 
In fact this follows from the previous comment, together with the fact that the equation of motion (\ref{EOMn}) is real up to 
the classical term involving the Chern-Simons level $k$.
\item The equations of motion are invariant under $\lambda_i\rightarrow-\lambda_i$, $\tilde\lambda_i\rightarrow -\tilde\lambda_i$.
\end{enumerate}
These are the same symmetries possessed by the ABJM theory with $n=1$  \cite{Herzog:2010hf}. 

The approximate equation of motion (\ref{approxEOM}) then becomes
\bea\label{crude}
-\frac{\partial F}{\partial \lambda_i} &\simeq& \left(1-\frac{2}{n+1}-\frac{n-1}{n+1}\right)\sum_{j=1}^N \sgn(\xi_i-\xi_j)~,\nn\\
&=& 0~,
\eea
where we have assumed that $\Re \lambda_i = \Re\tilde\lambda_i=\xi_i$, as in section \ref{saddle}. In the case at hand, 
this is also related to a \emph{symmetry}, namely $\zflip$. Since the action is invariant under this symmetry, 
it  is reasonable to expect that the same is true of
the saddle point solution. From the comments 1 and 2 above, this implies that 
$\tilde\lambda_i=\bar\lambda_i$, which in particular implies that $\Re \lambda_i = \Re\tilde\lambda_i$. 
The vanishing of the coefficient in (\ref{crude}) is of course precisely the single non-trivial NSVZ beta function relation, 
as expected from (\ref{NSVZ}).

\subsection{Evaluating the free energy}
\label{cont}

In this section we would like to evaluate the free energy in the M-theory large $N$ limit. 
In order to do so, it is convenient to use the $\zflip$ symmetry of the theory and action. 
From the comments made at the end of the previous subsection, this implies that
\bea
\lambda_i \, =\, N^\alpha x_i + \ii y_i~,\qquad \tilde\lambda_i \, = \, N^\alpha x_i - \ii y_i~.
\eea
In the continuum limit, this becomes $y(x)\equiv y^1(x)=-y^2(x)$. We may then 
use the general results in  (\ref{F2fourier}), (\ref{F2adjoint}), (\ref{F2gauge}) to write the total 
 \bea\label{F2Anfourier}
F_2 &=& \frac{4 N^{2-\alpha}}{\pi}\int_{x_1}^{x_2} (\rho(x))^2\diff x \sum_{m=1}^\infty 
\Bigg[ \frac{2}{m^3}\sin \frac{2\pi m}{n+1} \cos 2my 
+ \frac{1}{m^3}\sin \frac{2\pi m (n-1)}{n+1} \nn\\
&&+\frac{\pi}{m^2}+ 
\frac{2y}{m^2}  \sin \frac{2\pi m}{n+1} \sin 2my  - \frac{\pi (n-1)}{m^2(n+1)}
\cos \frac{2\pi m(n-1)}{n+1}\nn\\
&& - \frac{2\pi}{m^2(n+1)}\cos \frac{2\pi m}{n+1}\cos 2my\Bigg]~.
\eea
The sums in this expression are simple Fourier expansions, and it is elementary to 
sum them explicitly. For example, using
\begin{equation}\label{three}
\frac{1}{3}w^3 \, =\,  4\sum_{m=1}^\infty \frac{(-1)^m}{m^3}\sin mw + \frac{\pi^2}{3}w~, \qquad -\pi<w<\pi~,
\end{equation}
one easily shows that the first term in (\ref{F2fourier}) is
\begin{eqnarray}\label{bill}
\sum_{m=1}^\infty \frac{2}{m^3}\sin \frac{2\pi m}{n+1}\cos 2my& =& - \frac{1}{12}\left(\tfrac{\pi (n-1)}{n+1}+2y - 2\pi \epsilon\right)\left[
\left(\tfrac{(n-1)\pi}{n+1}+2y - 2\pi \epsilon\right)^2-\pi^2\right]\nn\\
&&   - \frac{1}{12}\left(\tfrac{\pi(n-1)}{n+1}-2y\right)\left[
\left(\tfrac{\pi(n-1)}{n+1}-2 y\right)^2-\pi^2\right]~.
\end{eqnarray}
The range on $y$ for the first term in (\ref{bill}) is 
\begin{eqnarray}\label{range1}
-\frac{\pi n}{n+1}<&y&<\frac{\pi}{n+1}~, \qquad ~~~~ \epsilon =0~,\nn \\ 
 \frac{\pi}{n+1}<&y&<\frac{\pi(n+2)}{n+1}~, \qquad \epsilon=1~,
\end{eqnarray}
 while for the second 
term it is 
\begin{equation}\label{range2}
-\frac{\pi}{n+1}<y<\frac{\pi n}{n+1}~.\end{equation} 
Notice that the range (\ref{range2}) has non-empty overlap with \emph{both} choices of range in (\ref{range1}). 
%In fact later we shall need to take $\epsilon=1$, which at the moment is not clear.

\begin{figure}[ht!]
\epsfxsize = 8cm
\centerline{\epsfbox{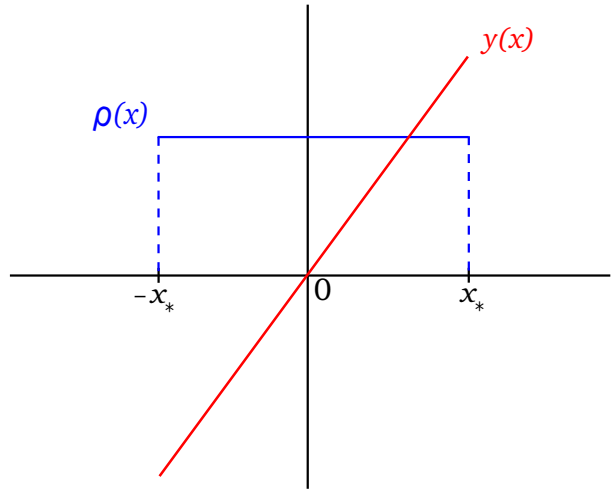}}
\caption{Sketch of $\rho(x)$, which is zero for $|x|>x_*$ and takes the constant 
value $\rho=\frac{1}{2x_*}$ for $|x|<x_*$. The function $y(x)$, shown in red, 
is linear in this latter region.}
\label{diagramAn}
\end{figure}

Using similar arguments it is straightforward to derive
\bea
F_2 &=& \frac{4N^{2-\alpha}}{\pi}\int_{x_1}^{x_2}(\rho(x))^2\diff x\Bigg\{\frac{2\pi}{(n+1)^3}\left[n^2\pi^2 -(n+1)^2 y^2\right]\nn\\
&& +  \frac{\pi^2\epsilon(\epsilon-1)}{3}\left[\frac{\pi}{n+1}\left[2\epsilon(n+1)+2-n\right]-3y\right]\Bigg\}~.
\eea
Remarkably, for $\epsilon=0$ \emph{or} $\epsilon=1$ this dramatically simplifies to the same expression
\bea\label{F2final}
F_2 &=& \frac{4N^{2-\alpha}}{\pi}\int_{x_1}^{x_2}(\rho(x))^2\diff x\frac{2\pi}{(n+1)^3}\left[n^2\pi^2 -(n+1)^2 y^2\right]~.
\eea
We see that the classical term (\ref{Fcl}) scales as $N^{1+\alpha}$ to leading order, while the 
one-loop term $F_2$ in (\ref{F2final}) scales as $N^{2-\alpha}$. In order to obtain 
non-trivial critical points in the large $N$ limit we thus need $1+\alpha=2-\alpha$, or 
$\alpha=\tfrac{1}{2}$. 
Altogether, the free energy is then, to leading order in the M-theory large $N$ limit, given by
\bea\label{freen}
F \, =\,  N^{3/2}\left[\frac{k}{\pi}\int \diff x x\rho(x) y(x) + \int \diff x (\rho(x))^2 h[y(x)]  - \frac{\mu}{2\pi}\left(\int \diff x \rho(x) -1\right)\right]~.
\eea
Here we have introduced a Lagrange multiplier $\mu$ for the density, and defined
\bea\label{hn}
h[y] &\equiv & \frac{8}{(n+1)^3}\left[n^2\pi^2 - (n+1)^2 y^2\right]~.
\eea
The Euler-Lagrange equations for a critical point are simply
\begin{eqnarray}
4\pi \rho(x)h[y(x)] &=& \mu - 2 k x y(x)~,\\
\pi \rho(x)h'[y(x)] &=& -kx~.
\end{eqnarray}
Computing $h' = -16 y/(n+1)$ one easily obtains the solution
\bea
\rho(x) &=& \frac{(n+1)^3 \mu}{32n^2\pi^3}~, \qquad y(x) \, =\,  \frac{2kn^2\pi^2 x}{(n+1)^2\mu}~.
\eea
Noting that the action is invariant under $x\leftrightarrow -x$, it follows that $x_1=-x_2$ and the constraint
equation is
\begin{equation}
\int_{-x_*}^{x_*} \rho(x)\diff x \, =\,  1 \quad \Rightarrow \quad \mu\, =\,  \frac{16 n^2\pi^3}{(n+1)^3x_*}~.
\end{equation}
In turn this gives
\bea
y &=& \frac{k(n+1)x_*}{8\pi}x~.
\eea
A sketch of these functions is shown in  Figure \ref{diagramAn}.

It is now a simple matter to substitute this back into the action and integrate, to obtain
\bea
F &=& N^{3/2}\left[\frac{4n^2\pi^2}{(n+1)^3 x_*}+\frac{k^2(n+1)}{48 \pi^2}x_*^3\right]~.
\eea
Finally, we must extremize over the endpoints of the eigenvalue distribution, so that $\diff F/\diff x_*=0$ implies 
\bea
x_* &=& \frac{2\pi}{n+1}\sqrt{\frac{2n}{k}}~, \qquad y(x_*) \, =\,  \frac{n\pi}{n+1}~.
\eea
%Looking back to  (\ref{range1}), (\ref{range2}), we see that this indeed lies in the range of validity of the Fourier 
%expansions we have made, provided we take $\epsilon=1$.
%
Looking back to (\ref{range1}), (\ref{range2}), we see that the full solution for $y(x)$
lies in the range of validity of the Fourier expansions we have made,
provided we take $\epsilon=1$ for the $y>\pi/(n+1)$ region in Figure \ref{diagramAn},
and $\epsilon=0$ for the $y<\pi/(n+1)$ region.
The leading saddle point free energy is hence
\bea
F &=& N^{3/2}k^{1/2}\frac{8\pi n^{3/2}}{3\sqrt{2}(n+1)^2}~.
\eea
For $n=1$ this is the ABJM result of \cite{Marino:2009jd, Drukker:2010nc, Herzog:2010hf}. We then see that
\begin{equation}
\frac{F(n=1)}{F(n)} = \sqrt{\frac{(n+1)^4}{16 n^3}}~,
\end{equation}
which is exactly the expected square root of the volume given by (\ref{vols}), (\ref{Voln}).

\subsection{The superconformal theory for $n>2$}
\label{nmorethantwo}

As discussed in section \ref{An}, for $n>2$ in fact the superpotential (\ref{superpotential}) cannot be 
marginal with $s\neq 0$, due to the Lichnerowicz/unitarity bound. In \cite{Martelli:2009ga} we therefore 
conjectured that for $n>2$ the coupling $s\Psi_I^{n+1}$ is \emph{irrelevant} in the IR, and thus 
one should set $s=0$ at the conformal fixed point. This then alters the above computation. The constraint
$\Delta[W]=2$ imposes
\bea\label{deltas}
\Da &=& 2(1-\Delta)~,
\eea
where we have set $\Delta\equiv \Db$.  Equation (\ref{deltas}) is also equivalent to (\ref{NSVZ}), which 
is the leading order saddle point equation. It is then straightforward to redo the computation of the previous 
section, with the \emph{weaker} condition (\ref{deltas}). One finds the free energy is still given by 
(\ref{freen}), but where the function $h[y]$ in (\ref{hn}) is 
\bea
h [y]  &=& 8(1-\Delta)(\pi^2\Delta^2 - y^2)~.
\eea
 Of course, setting $\Delta=\tfrac{n}{n+1}$ reproduces the function (\ref{hn}).
The rest of the computation proceeds in much the same way, and one finds the free energy
\bea\label{FDelta}
F(\Delta) &=& N^{3/2}k^{1/2}\frac{4\pi\sqrt{2\Delta^3(1-\Delta)}}{3}~.
\eea
As in \cite{Jafferis:2010un} this is a function of $\Delta$, which one should regard 
as a trial R-charge. Following the latter reference, extremizing $F$ with respect to 
$\Delta$ gives for the conformal field theory
\bea
\Delta &=& \Db \ = \ \frac{3}{4}~, \qquad \Da \ = \ \frac{1}{2}~.
\eea
Notice that this is formally equal to the previous result for $n=3$.

We may now compare with the dual gravity analysis. As shown in \cite{Martelli:2009ga}, 
the vacuum moduli space of the theory with $s=0$ is $\C\times \mathrm{Con}/\Z_k$. 
This certainly admits a Ricci-flat K\"ahler cone metric, namely the product 
of the flat metric on $\C$ times the conifold metric. More generally, one can compute 
the volume of a (singular) Sasakian metric on the link $Y$ using the results of 
\cite{Gauntlett:2006vf}. Realizing the conifold as the quadric hypersurface 
$\mathrm{Con}=\{z_1^2+z_2^2+z_3^2+z_4^2=0\}\subset \C^4$, we may compute the character 
of the action of $(\C^*)^2$ on $\C\times \mathrm{Con}$, where the first copy of $\C^*$ acts with 
weight one on $\C$ and the second $\C^*$ acts with weight one on each $z_i$, $i=1,\ldots,4$. 
In the notation of \cite{Gauntlett:2006vf}, this gives
\bea\label{character}
C(q_0,q,\C\times\mathrm{Con}) &=& \frac{1-q^2}{(1-q_0)(1-q)^4}~,
\eea
where $(q_0,q)$ are coordinates on $(\C^*)^2$. The volume, as a function 
of the Reeb vector field, is then given by setting $q_0=\ex^{-t\xi_0}$, 
$q=\ex^{-t\xi}$ and taking the limit of $t^{-4}$ times (\ref{character}) as 
$t\rightarrow 0$. This gives
\bea
\mathrm{Vol}(\xi_0,\xi) &=& \frac{2}{\xi_0\xi^3}\mathrm{Vol}(S^7)~,
\eea
where $\mathrm{Vol}(S^7)$ denotes the volume of the round metric on $S^7$.
One can easily check that the holomorphic $(4,0)$-form on $\C\times\mathrm{Con}$ has charge 4 
if and only if 
\bea
\xi_0 &=& 4 -2\xi~,
\eea
which is precisely the geometric analogue of (\ref{deltas}). Indeed, in the field theory 
$\Da=\xi_0/2$,
$\Delta=\xi/2$, which follows straightforwardly from the field theory 
description of the vacuum moduli space. We thus obtain the geometric formula
\bea
\mathrm{Vol}(\Delta)&=& \frac{1}{16\Delta^3(1-\Delta)}\mathrm{Vol}(S^7)~.
\eea
Remarkably, this precisely agrees with (\ref{FDelta}) and our conjecture (\ref{conjecture}); 
that is, the free energy is related to the Sasakian volume of $Y$, even before 
extremizing with respect to the trial R-charges. 

This result adds further support to the conjecture made in \cite{Martelli:2009ga} for the $n>2$ 
theories.  Notice that at this conformal fixed point $\Da=\tfrac{1}{2}$. Naively this contradicts 
the Lichnerowicz obstruction of \cite{Gauntlett:2006vf}, since it is stated there that $\Da=\tfrac{1}{2}$ 
can hold only for the round metric on $S^7$. However, the Lichnerowicz theorem of 
\cite{Gauntlett:2006vf} is here circumvented precisely because the link 
$Y$ is \emph{singular}. In fact certainly there is a Sasaki-Einstein metric on $Y$, which is 
known explicitly; but it has an $S^1$ locus of conical singularities. 

\section{The $U(N)^3$ SPP theory}\label{SPPsec}

In this section we study a different non-chiral Chern-Simons-matter theory, this time with 
$U(N)^3$ gauge group. 
We compute the partition function, as a function of the trial R-charges  \cite{Jafferis:2010un},
explicitly in the large $N$ limit, and verify that it satisfies our general Sasakian volume conjecture (\ref{conjecture}). 
 Note that although the Sasaki-Einstein metric is not known
in explicit form in this case, its existence is guaranteed by the results of \cite{FOW}.

\subsection{The quiver theory}

\begin{figure}[ht!]
\epsfxsize = 3.5cm
\centerline{\epsfbox{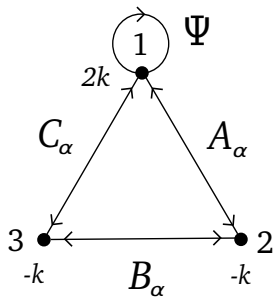}}
\caption{The SPP Chern-Simons quiver.}
\label{figSPP}
\end{figure}

The theory of interest is a $\mathcal{G}=U(N)_{2k}\times U(N)_{-k}\times U(N)_{-k}$ Chern-Simons-quiver theory, where the vector of Chern-Simons levels
is $(2k,-k,-k)$. Thus $G=3$. The matter content consists of the following bifundamental fields:
\bea
&& A_1: \mathbf{N}\otimes\overline{\mathbf{N}}\otimes 1~, \qquad A_2: \overline{\mathbf{N}}\otimes{\mathbf{N}}\otimes 1~,\nn\\
&& B_1: 1\otimes\mathbf{N}\otimes\overline{\mathbf{N}}~, \qquad B_2: 1\otimes \overline{\mathbf{N}}\otimes{\mathbf{N}}~,\nn\\
&& C_1: \mathbf{N}\otimes 1\otimes\overline{\mathbf{N}}~, \qquad C_2:  \overline{\mathbf{N}}\otimes 1\otimes{\mathbf{N}}~.
\eea
We also include an adjoint scalar field $\Psi$ for the first gauge group factor. The superpotential is
\bea\label{WSPP}
W&=& \mathrm{Tr}\, \left[\Psi\left(A_1A_2-C_1C_2\right)-A_2A_1B_1B_2+C_2C_1B_2B_1\right]~.
\eea
The quiver diagram is shown in Figure \ref{figSPP}.

As a four-dimensional $\mathcal{N}=1$ 
quiver gauge theory, this describes the low energy dynamics of $N$ D3-branes at the suspended pinch point (SPP) singularity \cite{Morrison:1998cs}. 
The latter is the (non-isolated) three-fold hypersurface singularity given by $\{x^2y=uv\}\subset \C^4$. 
From the general results of \cite{Martelli:2008si}, \cite{Aganagic:2009zk}, the corresponding  three-dimensional Chern-Simons-quiver theory 
is expected to describe $N$ M2 branes probing a related toric Calabi-Yau four-fold singularity. This Abelian moduli space 
was studied in both \cite{Aganagic:2009zk} and \cite{Hanany:2008fj}. 

Assuming the theory has a superconformal fixed point at which (\ref{WSPP}) is marginal, it follows from $\Delta[W]=2$ that
\bea\label{SPPconstraints}
&&2\, =\, 2\Delta[A] + 2\Delta[B]~, \qquad 2 \, = \, 2\Delta[C] + 2\Delta[B]~, \qquad 2 \, = \, 2\Delta[A] + \Delta[\Psi]~.
\eea
Here we have taken, by symmetry, $\Delta[A_1]=\Delta[A_2]=\Delta[A]$, {\it etc}. This leads to
\bea
&& \Delta[B]\, =\, \Delta~, \qquad \Delta[A]\, =\, \Delta[C]\, =\, (1-\Delta)~, \qquad \Delta[\Psi]\, = \, 2\Delta~.
\eea
In fact in this example the superpotential constraints (\ref{SPPconstraints}) are precisely equivalent 
to the vanishing of the four-dimensional NSVZ beta functions (\ref{NSVZ}). Recall that the latter 
are equivalent to the leading order saddle point equations for the partition function.

\subsection{The Sasakian volume function}

Until now it has not been possible to determine $\Delta$ at the superconformal fixed point using a 
purely field theory computation. However, one can determine $\Delta$ using the AdS/CFT correspondence 
together with the volume minimization of \cite{Martelli:2005tp}. In the latter reference it is 
shown how to uniquely determine the volume of a Sasaki-Einstein metric on the link of a 
toric Calabi-Yau singularity by minimizing a certain rational function. The latter is 
the volume of a general Sasakian metric as a function of the Reeb vector field $\xi$, which is 
holographically dual to the R-symmetry. This volume function is easily computed using the 
toric data of the Calabi-Yau singularity, and referring\footnote{In 
\cite{Hanany:2008fj} the authors have normalized the Reeb vector field so that 
the holomorphic $(4,0)$-form on the cone over $Y$ has charge 2, so that the superpotential 
then also has charge 2 (see the discussion after their equation (4.3)). In fact in order 
to obtain the correctly normalized volume this should be charge 4 \cite{Martelli:2005tp}, 
\cite{Martelli:2006yb}. 
Since the Sasakian volume function in this dimension is 
homogeneous degree $-4$, one should thus rescale all volumes in \cite{Hanany:2008fj}
by $2^{-4}=\tfrac{1}{16}$ in order to obtain the correct normalization. 
This factor will be crucial later when we compare to the large $N$ partition function for this model.} to equation (5.39) of \cite{Hanany:2008fj} we see that
in the present example (with $k=1$)
\bea\label{SPPvol}
\mathrm{Vol}(Y)[\Delta]
&=& \frac{4-3\Delta}{32\Delta(1-\Delta)^2(2-\Delta)^2}\mathrm{Vol}(S^7)~,
\eea
 where 
we have simplified somewhat  the expression given  in \cite{Hanany:2008fj}.
Here $\Delta$ is geometrically parametrizing the choice of Reeb vector field in the Sasakian metric 
on $Y$. However, this may be related to the field theoretic $\Delta$ using the 
correspondence between bifundamental fields in the Chern-Simons-quiver theory and 
M5 branes wrapped on (links of) certain toric divisors. More precisely, the AdS/CFT correspondence gives
\bea
\Delta[\Phi] &=& \frac{N\pi \mathrm{Vol}(\Sigma_\Phi)}{6\mathrm{Vol}(Y)}~,
\eea
where geometrically a bifundamental field $\Phi$ defines a line bundle over 
the moduli space, which is equivalent to a toric divisor $C(\Sigma_\Phi)$ with 
$\Sigma_\Phi\subset Y$ a codimension two subspace of $Y$.  Again, 
the volume of $\Sigma_\Phi$ may be computed using \cite{Martelli:2005tp}, and this 
identifies $\Delta$ in (\ref{SPPvol}) with the dimension $\Delta=\Delta[B]$. 

We note that in this case minimizing (\ref{SPPvol}) gives \cite{Hanany:2008fj} the conformal 
dimension\footnote{Note that although $\Delta< 1/2$, this does not violate
the unitarity bound because there is no gauge invariant operator with this dimension.}
\bea\label{cube}
\Delta &=& \frac{1}{18}\left[19-\frac{37}{\left(431-18\sqrt{417}\right)^{1/3}}-\left(431-18\sqrt{417}\right)^{1/3}\right]
\, \simeq\, 0.319~.
\eea

\subsection{Evaluating the free energy}
\label{eva}

Our goal in this section is to reproduce the geometric formula (\ref{SPPvol}), and hence 
following \cite{Jafferis:2010un} also (\ref{cube}), 
from a purely field theoretic computation. We simply apply the general results 
described in section \ref{gensection}. 

Recall that with $G=3$  gauge group factors we will have three sets of eigenvalues 
$\lambda^1_i$, $\lambda^2_i$, $\lambda^3_i$, where $i=1,\ldots,N$. Our choice 
of Chern-Simons levels in this example was determined by requiring the $\Z_2$ 
symmetry of the matter content and superpotential to extend to the whole 
Chern-Simons-matter theory. This $\Z_2$ acts by exchanging $A$ and $C$ fields and 
correspondingly the second and third $U(N)$ factors. Assuming that the saddle 
point solution is invariant under this $\Z_2$ symmetry of the theory, we may thus 
take 
\bea
&& y^1_i\, \equiv\, y_i~, \qquad y^2_i\, = \, y^3_i\, \equiv \, w_i~,
\eea
where recall that the eigenvalues are $\lambda^I_i=N^\alpha x_i+\ii y^I_i$. 
It it then straightforward to apply the general results (\ref{Fcl}), (\ref{F2fourier}), 
(\ref{F2adjoint}), (\ref{F2gauge}) to obtain the leading order partition function 
at large $N$. In particular, the one-loop contribution is
\bea
F_2 &=& \frac{2N^{2-\alpha}}{\pi}\int_{x_1}^{x_2}(\rho(x))^2\diff x \sum_{m=1}^\infty
\Bigg[\frac{2}{m^3}\sin 2\pi m(1-\Delta)-\frac{2\pi}{m^2}(1-\Delta)\cos 2\pi m(1-\Delta)\nn\\
&&+\frac{1}{m^3}\sin 2\pi m(1-2\Delta)-\frac{\pi}{m^2}(1-2\Delta)\cos 2\pi m(1-2\Delta)+\frac{3\pi}{m^2}\\
&&+\frac{4}{m^3}\sin 2\pi m\Delta \cos mu - \frac{4\pi}{m^2}\Delta\cos 2\pi m\Delta \cos mu + \frac{2}{m^2}u\sin 
2\pi m\Delta\sin mu\Bigg]~.\nn
\label{bigfur}
\eea
Here the first line comes from the $B$ fields, the second line from the adjoint $\Psi$ and the gauge sector (the last term), 
while the last line comes from the $A$ and $C$ fields. We have also introduced the quantity
\bea
u &\equiv & y-w~.
\eea
One can sum the Fourier series as before. Again, the classical and one-loop contributions are at the same order, thus 
leading to non-trivial solutions, only for $\alpha=\tfrac{1}{2}$. This leads to
\bea
F \,= \, N^{3/2}\Bigg[\frac{k}{\pi}\int \diff x\, x\rho(x)u(x) + \int \diff x\, (\rho(x))^2 h[u(x)] - \frac{\mu}{2\pi}\left(
\int\diff x\rho(x)-1\right)\Bigg]~,
\label{againf}
\eea
where
\bea
h[u]&=& 2\Delta\left[2\pi^2(\Delta-1)(\Delta-2)-u^2\right]~.
\label{niceh}
\eea
Importantly, as we discuss in Appendix \ref{interappe}, this is valid only for $u$ in the range
\bea
-2\pi(1-\Delta) \, <\,  u \, <\,   2\pi (1-\Delta) ~.
\eea

\begin{figure}[ht!]
\epsfxsize = 8cm
\centerline{\epsfbox{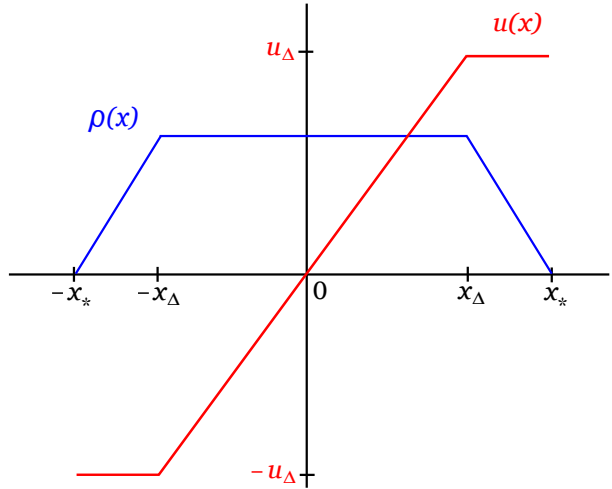}}
\caption{Sketch of the piecewise-linear functions $u(x)$, $\rho(x)$.}
\label{diagram}
\end{figure}

Solving the Euler-Lagrange  equations following from (\ref{againf}) we find the solution
\bea
\rho &=& \frac{\mu}{16\pi^3\Delta(1-\Delta)(2-\Delta)}~, \qquad u(x) \, = \, \frac{4\pi^2 k(1-\Delta)(2-\Delta)x}{\mu}~,
\eea
where without loss of generality let us take $k>0$. Notice that $\rho>0$ implies $\mu>0$.
Now, if we assume that this solution 
is valid in an interval $[-x_*,x_*] \subset [-x_\Delta, x_\Delta]$, where 
\bea
u(x_\Delta ) &\equiv & u_\Delta\, =\, 2\pi (1-\Delta)~,
\label{tap}
\eea 
we get a contradiction, since it then turns out that $u(x_*)>u_\Delta$. Hence this solution can be valid only in the interval $[-x_\Delta, x_\Delta]$, while in $[-x_*, -x_\Delta] \cup [x_\Delta, x_*]$ we necessarily have a different solution. Following  \cite{Herzog:2010hf},
in fact $u(x)$ is frozen to the  constant boundary value $u=u_\Delta$ in the interval $[x_\Delta, x_*]$, 
and correspondingly frozen to the other boundary value $u=-u_\Delta$ in the interval
$[-x_*, -x_\Delta]$. The resulting continuous, piecewise-linear function is shown in 
red in Figure \ref{diagram}.
Thus $F$ is not extremized with respect to $u(x)$ in the range $[-x_*, -x_\Delta] \cup [x_\Delta, x_*]$, but only with respect 
to $\rho (x)$ at fixed $u = \pm u_\Delta$. One finds the Euler-Lagrange equation solution
\bea
\rho (x) \, =\,  \frac{\mu - 4\pi k (1-\Delta) |x|}{16 \pi^3 \Delta^2 (1-\Delta)}~~~~~\mathrm{for}~~~~ x \in 
[-x_*, -x_\Delta] \cup [x_\Delta, x_*]~.
\eea
The value of $x_\Delta$ is easily determined from (\ref{tap}),
 while $x_*$ can be fixed by demanding that $\rho (x_*) =0$, giving 
\bea
x_\Delta\,  = \, \frac{\mu}{2\pi k (2-\Delta )}>0~, ~~~~~~~~~x_* \, =\,  \frac{\mu}{4\pi k (1-\Delta)}>0~,
\eea
respectively. 

We can then compute the on-shell $F$ as a function of $\mu$:
\bea
F & =  
& \frac{ N^{3/2}\mu^3 (4 - 3 \Delta) }{192 k \pi^5 (2-\Delta)^2 (1-\Delta)^2 \Delta }~.
\eea
Next,  $\mu$ may be determined by imposing the normalization of $\rho (x)$, giving
\bea
\mu^2 &=& \frac{ 64k \pi^4 \Delta   (1-\Delta)^2 (2-\Delta)^2}{(4 - 3 \Delta)} ~.
\eea
Finally, substituting this back into $F$ we obtain
\bea
F &=& N^{3/2} k^{1/2}  \frac{8\pi }{3} (2- \Delta  ) (1-\Delta ) \sqrt{\frac{\Delta}{  4- 3\Delta}}~,
\eea
which, remarkably, agrees with the conjectured relation (\ref{conjecture}) and the Sasakian volume (\ref{SPPvol}).

\section{Discussion}
\label{discus}

In this paper we have reproduced for the first time the  expected 
$N^{3/2}$ scaling of the number of degrees of freedom in ${\cal N}=2$  superconformal  field theories arising 
on  a large number $N$ of M2 branes,   where the conformal dimensions of matter fields  are 
\emph{different} from the canonical value $\Delta=\tfrac{1}{2}$. 
We have also reproduced the volumes of certain Sasaki-Einstein seven-manifolds, thereby providing non-trivial tests of some conjectured AdS$_4$/CFT$_3$ dualities \cite{Hanany:2008fj, Martelli:2009ga}. 
  More generally, we conjectured a relation between the large $N$
   limit of the partition function, interpreted as a function of trial R-charges,  and the volumes of Sasakian metrics on links of Calabi-Yau four-fold singularities.  
   
The method  \cite{Herzog:2010hf} that we have developed here applies 
to the general class of non-chiral quiver  theories. 
These  share some essential properties with ${\cal N}=3$ theories, albeit with 
superpotentials that are more arbitrary and not restricted to quartic interactions. Consequently, the scaling dimensions of the matter fields at the superconformal point are not restricted to be $\tfrac{1}{2}$.
However, we believe  that this is merely a technical difficulty and that an extension of the methods of this paper will allow one to 
tackle general ${\cal N}=2$ Chern-Simons-matter quivers. 
In particular, it will be important to justify more carefully some of the assumptions we and \cite{Herzog:2010hf} have made 
about the scaling of eigenvalues in the large $N$ limit. Although this ansatz reproduces the correct gravity results in the cases studied so far, 
and is supported by numerical results \cite{Herzog:2010hf}, clearly a more detailed understanding is desirable. This ansatz might need modification 
in more general classes of Chern-Simons-matter theories. 

It would be interesting to distill a simple general procedure for determining the
 R-charges of an arbitrary ${\cal N}=2$ Chern-Simons-quiver theory. This will presumably involve
 extracting an expression for the large $N$ free energy, as a function of the trial R-charges. 
 Recall that in $a$-maximization for four-dimensional theories one is given a simple 
 function (a cubic polynomial) of the trial R-charges to begin with. Again, the results of \cite{Martelli:2005tp, Martelli:2006yb}
 imply, via the AdS/CFT correspondence, that the square of the large $N$ free energy must be a polynomial function of the trial R-charges, at least in models with candidate gravity duals.
It would be very interesting to prove our conjecture, at least in sub-classes such as toric theories, using strategies analogous to those used in  \cite{Butti:2005vn, Eager:2010yu}.
Finally, it would be interesting to apply the matrix model techniques of \cite{Drukker:2010nc,Santamaria:2010dm} to the theories studied in this paper.

\subsection*{Acknowledgements}
We thank Nadav Drukker,  Nakwoo Kim, Elias Kiritsis and Vasilis Niarchos for discussions. 
D. M. would like to thank the Department of Physics of the University of Crete for allowing him to present 
the results of this paper prior to public release and for hospitality during its completion. 
D. M. is mainly supported by an EPSRC Advanced Fellowship EP/D07150X/3.
J. F. S. is  mainly supported by a Royal Society University Research Fellowship.

\appendix 

\section{Expansions}\label{expansions}

Recall the definition
\begin{equation}\label{ell2}
\ell(z) = -z \log\left(1-\ex^{2\pi\ii z}\right) + \frac{\ii}{2}\left[\pi z^2 + \frac{1}{\pi}\mathrm{Li}_2\left(\ex^{2\pi i z}\right)\right] - \frac{\ii\pi}{12}~.
\end{equation}
It is convenient to introduce the variable
\begin{equation}
\zeta = \ex^{2\pi \ii z} = \ex^{-2\pi \Im z } \left(\cos 2\pi \Re z + \ii \sin 2\pi \Re z\right) ~.
\end{equation}
Then for $\Im z>0$ we have the expansions 
\begin{eqnarray}
\log (1-\zeta) & = & - \sum_{m=1}^\infty \frac{\zeta^m}{m}~,\nn\\
\mathrm{Li}_2 (\zeta) & = & \sum_{m=1}^\infty \frac{\zeta^m}{m^2}~,
\end{eqnarray}
so that
\begin{eqnarray}
\ell(z) &=&  \frac{\ii\pi}{2}\left(z^2-\tfrac{1}{6}\right) + \sum_{m=1}^\infty \left(\frac{z}{m}+\frac{\ii}{2\pi m^2}\right)\ex^{2\pi\ii mz}~.
\label{expelle}
\end{eqnarray}
On the other hand, for $\Im z<0$ we have the expansions
\begin{eqnarray}
\log (1-\zeta) & = & \ii \pi + \log \zeta - \sum_{m=1}^\infty \frac{1}{m\zeta^m}~,\nn\\
\mathrm{Li}_2 (\zeta) & = & \frac{\pi^2}{3} - \ii \pi \log \zeta - \frac{1}{2} (\log\zeta)^2 -  \sum_{m=1}^\infty \frac{1}{m^2\zeta^m}~,
\end{eqnarray}
so that
\begin{eqnarray}
\ell(z) &=& - \frac{\ii\pi}{2}\left(z^2-\tfrac{1}{6}\right) + \sum_{m=1}^\infty \left(\frac{z}{m}-\frac{\ii}{2\pi m^2}\right)\ex^{-2\pi\ii mz}~.
\label{expelle2}
\end{eqnarray}
In summary, we have the following two series expansions for the function $\ell (z)$:
\begin{eqnarray}\label{ellpm}
\ell_+ (z) & = & \frac{\ii\pi}{2}\left(z^2-\tfrac{1}{6}\right) + \sum_{m=1}^\infty \left(\frac{z}{m}+\frac{\ii}{2\pi m^2}\right)\ex^{2\pi\ii mz}~, ~~~~~~~ \mathrm{for}~~ \Im z > 0~,  \nn\\
\ell_- (z) & = & - \frac{\ii\pi}{2}\left(z^2-\tfrac{1}{6}\right) + \sum_{m=1}^\infty \left(\frac{z}{m}-\frac{\ii}{2\pi m^2}\right)\ex^{-2\pi\ii mz}~,~~~ \mathrm{for}~~ \Im z < 0  ~.
\end{eqnarray}
We also note the following expansion. Writing $2\sinh \tfrac{w}{2}= \ex^{w/2}(1-\ex^{-w})$, we have
\begin{eqnarray}\label{wexp}
\log \, \left(2\sinh \tfrac{w}{2}\right) & = &  \frac{w}{2} - \sum_{m=1}^\infty \frac{1}{m}\ex^{-mw}~, ~~~~ \mathrm{for}~~ \Re w > 0~.
\end{eqnarray}

\section{More on the range of $u$ in section \ref{eva}}
\label{interappe}

By a  reasoning similar to that in section \ref{cont}, we find that the Fourier series in 
(\ref{bigfur}) may be resummed to the expression 
\bea\label{h}
h[u]&=& \frac{2}{3}\Bigg[2\pi^2\Big(3\Delta^3 - 9\Delta^2 + 3\Delta(2+p(1+p)+q(1+q)) -\\
&& (1+p+q)(p+q+2p^2+2q^2-2pq)\Big) + 3\pi u(q(1+q)-p(1+p)) -3\Delta u^2\Bigg]~,\nn
\eea
where {\it a priori} $p$ and $q$ are arbitrary integers. This is valid only if both conditions 
\bea
&&-\pi<\pi(1-2\Delta)+u+2\pi p<\pi~,\nn\\
&&-\pi<\pi(1-2\Delta)-u+2\pi q<\pi~,
\label{interv}
\eea
hold simultaneously. 
Notice that when $(p,q)$  take the values $(0,0)$, $(-1,0)$  and $(0,-1)$ the expression (\ref{h}) simplifies to
\bea
h[u] &=& 2\Delta\left[2\pi^2(\Delta-1)(\Delta-2)-u^2\right]~.
\label{nicehi}
\eea
Rewriting (\ref{interv}) as 
\bea
-2\pi(1-\Delta +p ) &<  ~u~ < & 2\pi (\Delta- p)~, \nn\\
-2\pi(\Delta -q ) & < ~u~ < & 2\pi (1-\Delta +q)~,
\label{interv3}
\eea
the analysis can be split into two cases, and we have
\bea
-2\pi(\Delta - q  ) &<  ~u~ < & 2\pi (\Delta- p)      ~~~~~~~~~~~\mathrm{for}~~ p+q+1> 2\Delta~, \label{aa}\\
-2\pi(1-\Delta +p  ) & < ~u~ < & 2\pi (1-\Delta +q)  ~~~~~~\mathrm{for}~~ p+q+1< 2\Delta ~,\label{bb}
\eea
where the two cases coincide when  $2\Delta = p+q +1$.  
In order for the interval in (\ref{aa}) to be as large as possible, the integers $p$ and $q$ should be ``as negative as possible''. However, if for example $p=0, q=-1$ we have $\Delta <0$, hence they cannot be negative. The largest interval is thus obtained for $p=q=0$ and we have
\bea
-2\pi\Delta\,  <\,   u \, <\,   2\pi \Delta   ~~~~~~~~~~~\mathrm{for}~~ p=0, \, q=0~.
\eea
 Similarly, in order for the interval in (\ref{bb})
 to be as large as possible, the integers $p$ and $q$ should be as large (and positive) as possible. 
At this point, let us take a short-cut and use the information  that $\Delta\sim 0.32<1/2$ for the superconformal fixed point.   Then notice that if $q=p=0$ we would  
have $\Delta >1/2 $, so at least one of $p$, $q$ must be negative and we have the two solutions 
\bea
2\pi\Delta  &<  ~u~ < & 2\pi (1- \Delta)      ~~~~~~~~\mathrm{for}~~ p=-1,\, q=0~,\\
-2\pi(1-\Delta   ) & < ~u~ < & -2\pi \Delta   ~~~~~~~~~~~~~\mathrm{for}~~ p =0,\, q=-1~.
\eea
Remarkably, in all cases the function $h[u]$ simplifies to the expression (\ref{nicehi}). Therefore, putting everything together, we conclude that the range of validity (at least when $\Delta <1/2$) of (\ref{nicehi}) is 
\bea
-2\pi(1-\Delta) \, <\,   u \, <\,   2\pi (1-\Delta) ~,
\eea
as claimed in section \ref{eva}.

\end{document}